\documentclass[letterpaper,twocolumn,english,aps,prl,superscriptaddress]{revtex4-1}
\usepackage[T1]{fontenc}
\usepackage{color}
\usepackage{babel}
\usepackage{amsmath}
\usepackage{amssymb}
\usepackage{graphicx}
\usepackage[unicode=true,pdfusetitle,
 bookmarks=true,bookmarksnumbered=true,bookmarksopen=true,bookmarksopenlevel=3,
 breaklinks=true,pdfborder={0 0 1},backref=false,colorlinks=true]
 {hyperref}
\hypersetup{
 pdfborderstyle=}

\makeatletter

\pdfpageheight\paperheight
\pdfpagewidth\paperwidth

\usepackage{babel}

\usepackage{mathalpha}
\usepackage{xcolor}
\usepackage{simpler-wick}
\usepackage{footnote}
\usepackage{url}

\def\e{\textrm{e}}
\def\i{\textrm{i}}
\def\td{\textrm{d}}

\makeatother

\begin{document}
\title{Coherent Axion Production through Laser Crystal Interaction}

\author{Zhan Bai}
\email{baizhan@siom.ac.cn}
\affiliation{State Key Laboratory of High Field Laser Physics, Shanghai Institute
of Optics and Fine Mechanics, Chinese Academy of Sciences}
\affiliation{CAS Center for Excellence in Ultra-intense Laser Science}

\author{Xiangyan An}
\affiliation{Tsung-Dao Lee Institute, Shanghai Jiao Tong University}

\author{Yuqi Chen}
\affiliation{Institute of Theoretical Physics, Chinese Academy of Sciences}

\author{Baifei Shen}
\affiliation{Department of Physics, Shanghai Normal University}

\author{Ruxin Li}
\affiliation{Shanghai Tech University}

\author{Liangliang Ji}
\email{jill@siom.ac.cn}
\affiliation{State Key Laboratory of High Field Laser Physics, Shanghai Institute
of Optics and Fine Mechanics, Chinese Academy of Sciences}
\affiliation{CAS Center for Excellence in Ultra-intense Laser Science}

\date{\today}
\begin{abstract}
We investigate the interaction between an optical laser and an ionic
crystal and reveal coherent emission of axions through phase-match
between laser and axion fields. Such emission is further enhanced
by stacking thin crystal layers of half-wavelength thickness. Based
on these findings, we propose a novel method for generating and detecting
axions in terrestrial experiments, 
achieving up to a two-order-of-magnitude increase in transition probability compared to light-shining-through-wall
(LSW) experiments with the same interaction region size. For an experimental
length of 10 meters, this setup could lower the exclusion limit
to $g_{a\gamma\gamma}\gtrsim1.32\times10^{-11}\text{GeV}^{-1}$ with
currently available laser technologies. 
\end{abstract}
\maketitle
\textbf{\emph{Introduction}}\emph{. } Dark matter, a cornerstone of
physics beyond the Standard Model, remains elusive (for review, see e.g.
Ref.~\citet{Bertone:2004pz}). Among its numerous candidates, axion
is particularly promising. It was originally proposed to explain why
the charge-parity violation term is extremely small in quantum chromodynamics
(i.e. the strong CP problem)~\citep{Peccei:1977hh,Peccei:1977ur},
but was later found to be a perfect candidate for dark matter\citep{Dine:1982ah,Abbott:1982af,Preskill:1982cy}.
This dual relevance has spurred extensive experimental efforts, targeting
both axions and axion-like particles.

Axion detection often relies on its coupling to electromagnetic fields.
For instance, the CAST experiment detects solar axions by converting
them into photons in a strong magnetic field\citep{CAST:2007jps,CAST:2017uph}.
Similarly, CDMS experiments use germanium crystals, where axions are
converted to photons through electric fields in atoms\citep{CDMS:2009fba,SuperCDMS:2022kse},
with Bragg condition-induced coherence enhancing signals\citep{Creswick:1997pg}.

Terrestrial experiments, such as light-shining-through-wall (LSW)
setups, attempt to produce axions using the reverse of these mechanisms.
Lasers interacting with magnetic fields produce axions, which then
cross a wall and are reconverted to photons on the other side\citep{OSQAR:2015qdv,Kozlowski:2024jzm,Ehret:2010mh}.
Other experiments, such as PVLAS\citep{DellaValle:2015xxa,Ejlli:2020yhk},
use similar axion generation approaches. Enhancing axion production
rates typically requires stronger or longer magnetic fields, but this
approach is costly and technically challenging. Current LSW experiments
use a magnetic field with $BL=129\text{\,Tm}$\citep{OSQAR:2015qdv}
and plans to increase to $BL=562\text{\,Tm}$\citep{Kozlowski:2024jzm}.

To improve axion production, the strong electric fields inside crystals
are harnessed, for instance, with X-rays interacting with crystals.
Axions can be produced coherently when the Bragg condition is satisfied.
However, absorption of X-ray photons by crystals limits the interaction
distance to millimeter scale or even less \citep{Buchmuller:1989rb,Henke:1993eda,Yamaji:2017pep,Halliday:2024lca},
far shorter than that in typical LSW setups ($\gtrsim10\text{\,m}$).

Inspired by these studies, we find a new approach to overcome these
challenges and significantly boost the production rate of axions:
optical laser interaction with transparent ionic crystals. Compared
to X-rays, optical lasers possess higher photon densities and enable
much longer propagation distance in crystals. Yet, in covalent crystals,
Coulomb fields are shielded by electron clouds, making it highly localized
within atoms. Such fields are difficult to be sensed by optical lasers
with micro-meter wavelength. We propose employing ionic crystals,
wherein the Coulomb fields are much more widespread. Using this combination,
we demonstrate that aligning light at specific angles and stacking
thin crystal layers leads to a novel coherence mechanism, which is
different from the Bragg-type enhancement. This approach significantly
boosts axion production, offering a pathway to tighter constraints
on axion coupling constants.

\textbf{\emph{Coherent Axion Production in Medium. }} When a laser
interacts with a point-like charge, the magnetic field of the laser
is coupled with the Coulomb field, inducing overall non-zero $\boldsymbol{E}\cdot\boldsymbol{B}$,
and axion field is excited according to wave equation~\citep{CAST:2007jps}
$\left(\partial_{t}^{2}-\nabla^{2}+m_{a}^{2}\right)a=g_{a\gamma\gamma}\boldsymbol{E}\cdot\boldsymbol{B}$,
where $a$ is the axion field, $m_{a}$ is the axion mass, $g_{a\gamma\gamma}$
is the coupling constant, and $\boldsymbol{E}$ and $\boldsymbol{B}$
are the electric and magnetic field, respectively. Once the Coulomb
fields are arranged in a periodic manner with scale length much smaller
than the laser wavelength, the contribution from each are coherently
superposed. In this case the axion production number is integrated\citep{Peskin:QFT}
\begin{equation}
N_{a}=  \int\frac{\td^{3}\boldsymbol{k}_{a}}{\left(2\pi\right)^{3}}\frac{1}{2E_{a}}\left|\tilde{j}\left(k_{a}^{0},\boldsymbol{k}_{a}\right)\right|_{k_{a}^{0}=E_{a}}^{2}.\label{eq:ScalarNumberFromClassicalSource}
\end{equation}
Here $\boldsymbol{k}_{a}$ is the momentum of axion, $E_{a}=\sqrt{\boldsymbol{k}_{a}^{2}+m_{a}^{2}}$
is the energy of axion, and $\tilde{j}\left(k_{a}^{0},\boldsymbol{k}_{a}\right)$
is the Fourier transformation of the source term $j\left(t,\boldsymbol{r}\right)\equiv g_{a\gamma\gamma}\boldsymbol{E}\cdot\boldsymbol{B}$.
We consider a linearly polarized laser whose magnetic field is $\boldsymbol{B}=\boldsymbol{B}_{0}\cos\left(\omega t-\boldsymbol{k}_{L}\cdot\boldsymbol{r}\right)$,
with $\boldsymbol{k}_{L}$ being the wave vector, and $\omega$ the
circular frequency. 

For many ions of charges $q_{s}$ at locations $\boldsymbol{r}_{s}$,
represented by the electric field $\boldsymbol{E}=\sum_{s}q_{s}\left(\boldsymbol{r}-\boldsymbol{r}_{s}\right)/\left(4\pi\left|\boldsymbol{r}-\boldsymbol{r}_{s}\right|^{3}\right)$,
the conversion probability defined as $P=N_{a}/N_{\gamma}$ ($N_{\gamma}$
is the laser photon number), is thus (for detailed derivation, see
Appendix.~\ref{sec:ClassicalFieldApproach}): 
\begin{equation}
P_{\text{laser}\rightarrow a}=  \frac{g_{a\gamma\gamma}^{2}}{S}\int\td\Omega\frac{\td P_{\text{single}}}{\td\Omega}\left|\mathcal{T}\right|^{2},\mathcal{T}\equiv\sum_{s}q_{s}\e^{-\i\Delta\boldsymbol{k}\cdot\boldsymbol{r}_{s}}\label{eq:TransitionProbability:Linear}
\end{equation}
where $S$ is the laser focal area, $\Omega$ is the solid angle,
$\td P_{\text{single}}/\td\Omega$ is the differential conversion
probability for a single charge at the origin, and $\Delta\boldsymbol{k}=\boldsymbol{k}_{a}-\boldsymbol{k}_{L}$
is the momentum transfer. It should be noted that while the $\boldsymbol{E}$
field inside the ions can be different from point charge, the corresponding
modification is negligible when the ion radius is much shorter than
laser wavelength (see Appendix.~\ref{sec:In-Atom}). The superposition
of contributions from all ions are contained in the translation term
$\mathcal{T}$. 

Eq.(\ref{eq:TransitionProbability:Linear}) applies for charges at
arbitrary positions. For a regularly placed crystal lattice, the translation
term can be split into two parts: $\mathcal{T}\equiv\mathcal{T}_{\text{cell}}\cdot\mathcal{T}_{\text{lat}}$,
defined as: 
\begin{equation}
\mathcal{T}_{\text{cell}}\equiv\sum_{c}q_{c}\e^{-\i\Delta\boldsymbol{k}\cdot\delta\boldsymbol{r}_{c}},\qquad\mathcal{T}_{\text{lat}}\equiv\sum_{l}\e^{-\i\Delta\boldsymbol{k}\cdot\boldsymbol{r}_{l}}\label{eq:T_cell_lat}
\end{equation}
where $\delta\boldsymbol{r}_{c}$'s are the relative positions of
particle $c$ in one cell, and $\boldsymbol{r}_{l}$'s are the reference
coordinates for unit cells on lattice. From Eq.(\ref{eq:T_cell_lat}),
it is clear that for optical lasers if the particles are all charge
neutral atoms, $\mathcal{T}_{\text{cell}}$ vanishes and no axion
signal is produced. Consider a simplest cell with two opposite charges
$\pm q$ separating a distance $\delta\boldsymbol{r}$, the dipole
moment leads to $\mathcal{T}_{\text{cell}}\approx\i q\Delta\boldsymbol{k}\cdot\delta\boldsymbol{r}$,
which is of order $\mathcal{O}(\omega|\delta\boldsymbol{r}|)$. Unlike
in crystals of neutral atoms, such dipoles exist in ionic crystals
where positive and negative charges are arranged in an interleaved
pattern, leading to finite translation term for each cell $\mathcal{T}_{\text{cell}}$.
Contributions from these cells at various locations are superposed
following their phases in axion production, represented by the $\mathcal{T}_{\text{lat}}$
term.

\begin{figure*}
\begin{centering}
\includegraphics[width=0.95\textwidth]{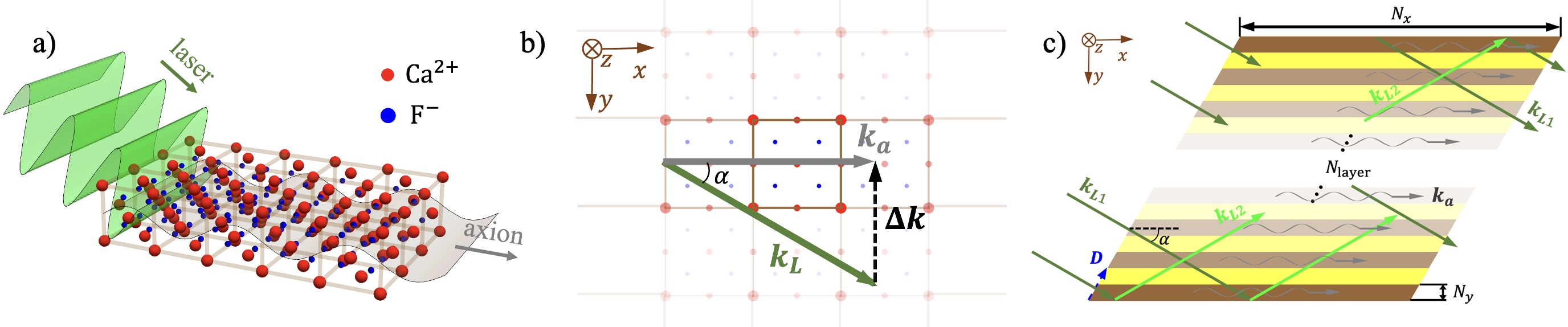} 
\par\end{centering}
\vspace{-10pt}
 \caption{ \emph{Panel (a)}: A schematic figure for the coherent production
of axion in laser-crystal interactions. The green and gray surfaces
indicate the incoming laser and the outcoming axion fields, respectively.
The red and blue spheres indicate positive and negative ions, respectively.
Each brown cubic represents an unit cell. \emph{Panel (b)}: A side
view of panel (a). The brown squares indicate the crystal cells. $\boldsymbol{k}_{L}$
and $\boldsymbol{k}_{a}$ are the wave vectors for incoming light
and outcoming axion, respectively. \emph{Panel (c)}: A schematic figure
for the stacking of thin crystal films. The brown layers represent
the ionic crystal materials. Every brown layer has $N_{x}\times N_{y}\times N_{z}$
lattice cells, and each layer shifts a constant $\boldsymbol{D}$
in the direction perpendicular to the laser direction $\boldsymbol{k}_{L1}=\boldsymbol{k}_{L}$.
In the uppermost and lowermost layers, we present a schematic figure
for the reflection. Mention that $\boldsymbol{k}_{L1}$ is the wave
vector for the incident laser and also the laser reflected by the
upper boundary, and $\boldsymbol{k}_{L2}$ is the wave reflected by
the lower boundary. The yellow areas are the supporting material.\label{fig:CoherentDirection-DispersionCrystal}}
\vspace{-10pt}
 
\end{figure*}

We consider an ionic crystal-calcium fluoride ($\text{CaF}_{2}$),
as an illustration for coherent axion emission, although our discussion
applies for all transparent ionic crystals. The $\text{CaF}_{2}$
crystal have face-centered-cubic(FCC) structure~\citep{CaF2:2024Wikipedia:nov},
where $\text{Ca}^{2+}$ are located at the corners and the center
of each face of the cube, and $\text{F}^{-}$ occupy all the tetrahedral
voids (holes) within the lattice. The interaction of laser pulse with
the crystal structure is shown in Fig.\ref{fig:CoherentDirection-DispersionCrystal}(a).
In one cell, there are 4 $\text{Ca}^{2+}$ and 8 $\text{F}^{-}$ ions.
The lattice constant is $d=0.5451\text{nm}$, and the refractive index
is $n=1.43$ for laser wavelength $\lambda=2\pi/\omega=1064\text{nm}$\citep{Polyanskiy:RefractiveIndex}.

As shown in Fig.\ref{fig:CoherentDirection-DispersionCrystal}(b),
the laser injects into a crystal with an inclination angle $\alpha$.
From Eq.(\ref{eq:T_cell_lat}), we see that if $\Delta\boldsymbol{k}$
is parallel to the transverse $y$-axis, cells with the same $y$-coordinates
have the same phases, which contribute to a coherent enhancement.
As the axion field travesl along the longitudinal $x$-axis, $\boldsymbol{k}_{L}$,
$\boldsymbol{k}_{a}$ and $\Delta\boldsymbol{k}$ together form a
phase-match condition, as shown in Fig.~\ref{fig:CoherentDirection-DispersionCrystal}(b).

When axion mass $m_{a}\ll\omega$, the momentum of the outgoing axion
satisfies $\left|\boldsymbol{k}_{a}\right|\approx\omega=\left|\boldsymbol{k}_{L}\right|/n$.
The phase-match requires that $\alpha=\arccos\frac{1}{n}$, exactly
the angle of full reflection. We can then complete the summation in
$\mathcal{T}_{\text{lat}}$ and its maximum value is: 
\begin{align}
\left|\mathcal{T}_{\text{lat}}^{\text{max}}\right|^{2}= & N_{x}^{2}\frac{1-\cos\left(N_{y}\omega d\tan\alpha\right)}{1-\cos\left(\omega d\tan\alpha\right)}N_{z}^{2},\label{eq:Translation:Lattice}
\end{align}
where $N_{x}$, $N_{y}$ and $N_{z}$ is the number of unit cells
along $x$, $y$ and $z$ direction, respectively. One notices coherent
enhancement for large $N_{x}$ and $N_{z}$, but periodically oscillates
with $N_{y}$. It is because $\Delta\boldsymbol{k}$ is parallel to
$y$-axis and as the phase $\Delta\boldsymbol{k}\cdot\boldsymbol{r}_{l}$
changes along the $y$ direction, the signal undergoes coherent enhancement
and coherent annihilation periodically. 

\textbf{\emph{Layer Structure}}. According to Eq.(\ref{eq:Translation:Lattice}),
the contribution from $y$ direction reaches its maximum when $N_{y}^{\text{max}}\omega d\tan\alpha=(2M+1)\pi$
for any integer $M$. However a larger $M$ leads to a thicker crystal
and larger focal area, which reduces conversion probability, as shown
in Eq.(\ref{eq:TransitionProbability:Linear}). We will therefore
take $M=1$ for optimized axion conversion rate. For $\text{CaF}_{2}$
this means $N_{y}^{\text{max}}\approx950$ and $L_{y}^{\text{max}}=N_{y}^{\text{max}}d\approx518\text{nm}$,
essentially a thin film. Further increase in $N_{y}$ leads to the
coherent annihilation and then enhancement and so on. The signals
will oscillate according to Eq.(\ref{eq:Translation:Lattice}). 

It will be ideal if the annihilation phase $(2m-1)\pi\le N_{y}\omega d\tan\alpha\le2m\pi$
for any integer $m$, are inactivated to maintain continuous grow
for large $N_{y}$. To do this, we propose stack multiple crystal
layers along the $y$-axis, as is shown in Fig.\ref{fig:CoherentDirection-DispersionCrystal}(c).
Each layer is placed with a shift $\boldsymbol{D}$. They contribute
to the translation term with a phase factor $\exp\left(-\i n\Delta\boldsymbol{k}\cdot\boldsymbol{D}\right)$
for the $n$-th layer. For best coherence, we require that $\Delta\boldsymbol{k}\cdot\boldsymbol{D}=2N\pi$,
where $N$ is an arbitrary integer. Similar to the choice of $M$,
we take $N=1$ for smallest focal area and largest conversion probability.
We then have $D_{y}=(2\pi/\omega)\cot\alpha$. We also want each layer
on the same phase front of the laser, i.e. $\boldsymbol{D}\perp\boldsymbol{k}_{L}$,
so we have $D_{x}=D_{y}/\cot\alpha=2\pi/\omega$.

The spaces between displaced thin ionic crystal layers should be filled
with other transparent materials for stability of the structure. Those
materials should be atom crystals that are inactive for axion production
to avoid coherent annihilation (see Appendix.~\ref{sec:In-Atom}).
In order for the light to propagate between layers, the supporting
layers should have similar refractive index. For example, the $\textrm{SiO}_{2}$
crystal has $n=1.45$\citep{Polyanskiy:RefractiveIndex}, which is
close to $\text{CaF}_{2}$.

In this staked layers, since our choice of propagating direction $\alpha=\arccos1/n$
is the critical angle for total reflection, the laser keeps reflecting
while propagating along $x$ direction, on the uppermost and lowermost
layer surface, as indicated in Fig.\ref{fig:CoherentDirection-DispersionCrystal}(c).
This is equivalent to a rectangular
waveguide working in transverse electronic (TE) mode.
We can then
treat the wave function as two propagating plane waves, with wave
vectors $\boldsymbol{k}_{L1}=n\omega\left(\cos\alpha,\sin\alpha,0\right)$
and $\boldsymbol{k}_{L2}=n\omega\left(\cos\alpha,-\sin\alpha,0\right)$.
Substituting into Eq.(\ref{eq:ScalarNumberFromClassicalSource}),
we can derive the conversion probability. 

\begin{figure}
\begin{centering}
\includegraphics[width=0.45\textwidth]{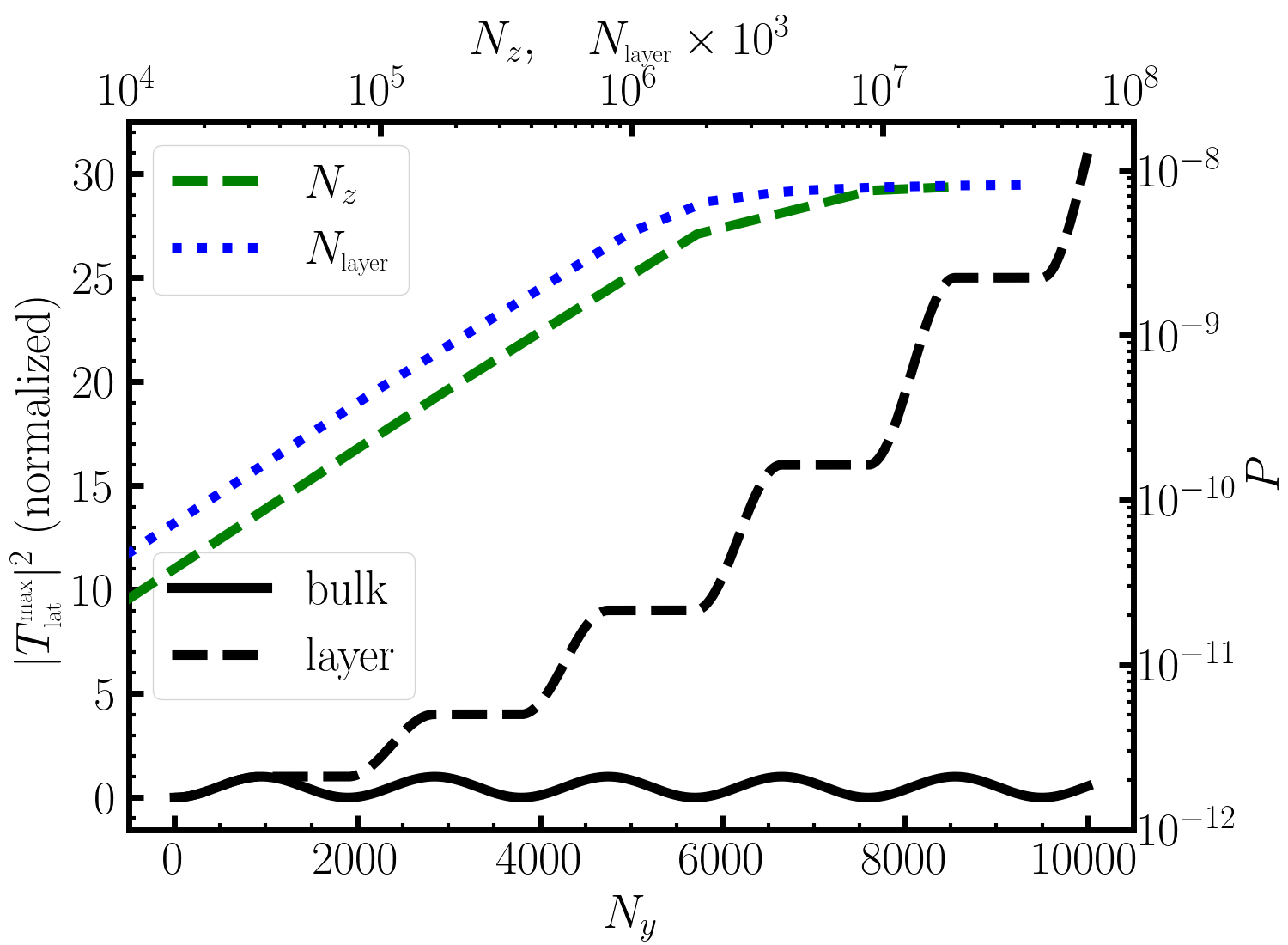} 
\par\end{centering}
\vspace{-5pt}
 \caption{ Scaling behavior with the increase of crystal sizes. Black lines
(left and bottom axes) show $|\mathcal{T}_{\text{lat}}|^{2}$ versus
$N_{y}$. The solid line represents oscillations in a complete bulk
crystal (Eq.(\ref{eq:Translation:Lattice})), while the dashed line
shows growth when layer structure is used. The left $y$-axis is normalized
so that the maximum value for the black solid line is $1$. Colored
lines (top and right axes) indicate conversion probability growth
with increasing $N_{z}$ and $N_{\text{layer}}$. \label{fig:Scaling}}
\vspace{-20pt}
 
\end{figure}

The detailed derivation and explicit formula for the conversion probability
can be found in Appendix.~\ref{sec:Two-Plane-Waves} and Eq.(\ref{eqn:TwoWaveProbability}).
Here we only analyze the scaling behavior qualitatively. Eq.(\ref{eq:Translation:Lattice})
presents the maximum value of $|\mathcal{T}_{\text{lat}}|^{2}$ when
the phase-match condition is satisfied. It oscillates periodically
if $N_{y}$ increases monotonically, as shown in Fig.~\ref{fig:Scaling}
with the black solid line. However, if we keep only the increasing
region and inactivate the decreasing region, i.e., if we use layer
structure, the $|\mathcal{T}_{\text{lat}}|^{2}$ will increase continuously.
The summation of all layers will contribute a factor $\left[\sum_{n}\exp\left(-\i n\Delta\boldsymbol{k}\cdot\boldsymbol{D}\right)\right]^{2}\approx N_{\text{layer}}^{2}$
to $|\mathcal{T}_{\text{lat}}^{\text{max}}|^{2}$, where $N_{\text{layer}}$
is the total number of layers, as shown in Fig.~\ref{fig:Scaling}
with the black dashed line.

$|\mathcal{T}_{\text{lat}}|^{2}$ is strongly peaked near phase-match
condition, which leads to extremely collimated axion emission with
divergence angle $\Delta\Omega$. The integration in Eq.(\ref{eq:TransitionProbability:Linear})
is then approximately $P\propto\frac{1}{S}N_{\text{layer}}^{2}\left|\mathcal{T}_{\text{cell}}\mathcal{T}_{\text{lat}}^{\text{max}}\right|^{2}\Delta\Omega$.
The divergence angle, $\Delta\Omega$, is proportional to $\lambda^{2}/S$.
Since $S\propto N_{\text{layer}}N_{z}$, we can see the $N_{\text{layer}}^{2}$
dependence and $N_{z}^{2}$ dependence in $|\textrm{T}_{\text{lat}}^{\text{max}}|^{2}$
are canceled by the focal area $S^{2}$. Therefore, the conversion
probability only scales as $N_{x}^{2}$, and does not rely on $N_{z}$
and $N_{\text{layer}}$, as long as the crystal is large enough and
enter the scaling region. 

We numerically calculate the conversion probability for $g_{a\gamma\gamma}=10^{-7}\text{GeV}^{-1}$
and $m_{a}=10^{-6}\text{eV}$, with different crystal sizes, as shown
in Fig.~\ref{fig:Scaling} with colored lines. $N_{x}=1.83\times10^{10}$
and $N_{y}=950$ is fixed for these two lines, and the conversion
probability clearly increases when $N_{z}$ and $N_{\text{layer}}$
are small, but saturates for $N_{z}\gtrsim10^{7}$ and $N_{\text{layer}}\gtrsim3\times10^{3}$. 

Therefore, we consider $N_{x}=1.83\times10^{9}$, $N_{y}=950$, $N_{z}=9.17\times10^{6}$,
and $N_{\text{layer}}=4.8\times10^{3}$, so that the target is a thin
rod with size $1\text{m}\times5\text{mm}\times5\text{mm}$. The conversion
probability is $P_{\text{laser}\rightarrow a}^{1\text{m}}=8.53\times10^{-11}$.
If we increase $N_{x}$ to $1.83\times10^{10}$, i.e. consider a 10-meter-long
rod, the conversion probability will be $P_{\text{laser}\rightarrow a}^{10\text{m}}=7.58\times10^{-9}$.
For comparison, in LSW experiment\citep{OSQAR:2015qdv}, the conversion
probability is $P^{\text{OSQAR}}=4.06\times10^{-11}$ for the same
axion mass and coupling, with magnetic field of length $L=14.3\text{m}$.
Therefore, if the length scale of the crystal is of the same order
as the magnetic field length for LSW experiment, the conversion probability
can be two order of magnitude higher.

\textbf{\emph{Reconversion}}. As we have stated, the axion beam is
highly collimated. For the crystal size we use, the divergence is
around $\Delta\theta\lesssim3\times10^{-5}\pi$. This highly directional
axion beam is favorable for reconversion into light for detection.
By injecting the axion into another crystal, they will convert back
to light by interacting with the Coulomb field of the ions. For crystals
inside a rectangular waveguide, TE mode will be excited. The 
physical picture is the inverse of axion production, and phase match
condition is identical, so the layer structure is required. The detailed
derivation is shown in Appendix.\ref{sec:Reconversion-in-Waveguide}
and Eq.(\ref{eq:P_re:waveguide}).

We again consider $g_{a\gamma\gamma}=10^{-7}\text{GeV}^{-1}$ and
$m_{a}=10^{-6}\text{eV}$ for comparison. For a $\text{CaF}_{2}$
stick with size $1\text{m}\times5\text{mm}\times5\text{mm}$, i.e.
with $N_{x}=1.83\times10^{9}$, $N_{y}=950$, $N_{z}=9.17\times10^{6}$,
and $N_{\text{layer}}=4.8\times10^{3}$, the conversion probability
is $P_{a\rightarrow\gamma}^{\text{1m}}=1.85\times10^{-10}$. If we
further increase the length to $10\text{m}$, we have $P_{a\rightarrow\gamma}^{\text{10m}}=1.85\times10^{-8}$.

\begin{figure}
\begin{centering}
\includegraphics[width=0.45\textwidth]{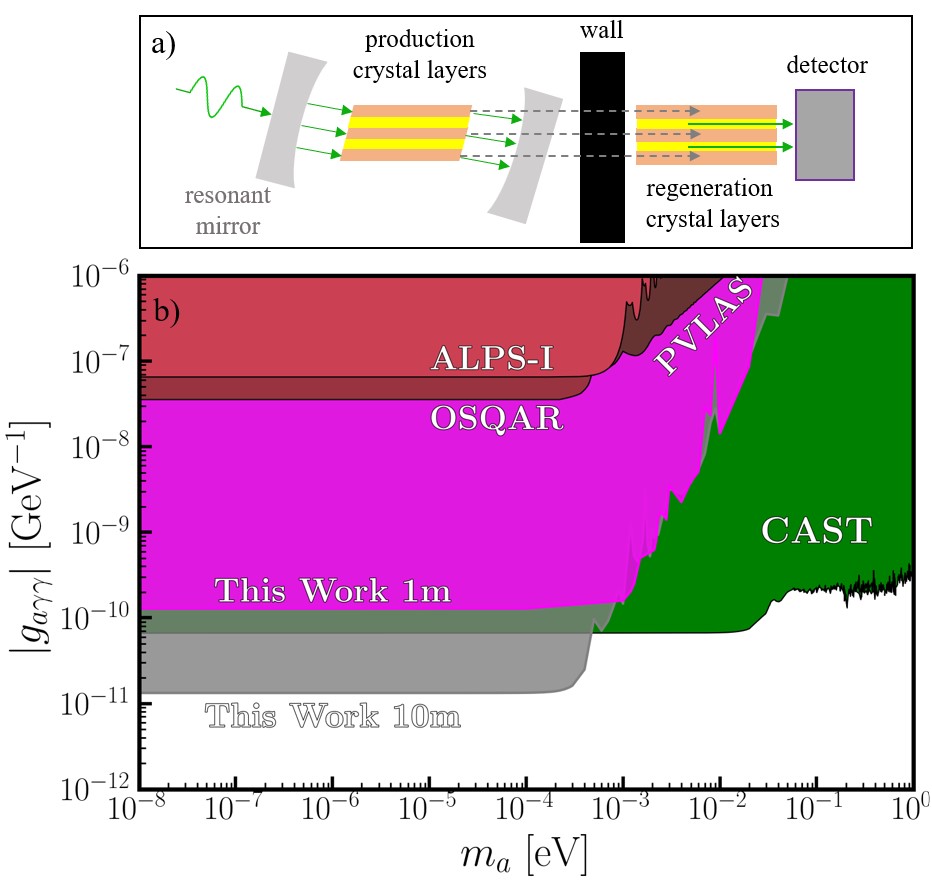} 
\par\end{centering}
\vspace{-10pt}
\caption{ \emph{Upper panel:}Schematic figures for the experimental setup.
\emph{Lower panel:} The exclusion line for our proposal. The shaded
pink region is excluded using $1\text{m}$ conversion and reconversion
length, and the shaded gray region is excluded using $10\text{m}$
conversion and reconversion region. The exclusion lines from LSW\citep{OSQAR:2015qdv,Ehret:2010mh},
PVLAS\citep{DellaValle:2015xxa} and CAST\citep{CAST:2017uph} experiments
are also given as comparison. The figure is plotted using Axion Limits\citep{AxionLimits}.\label{fig:ExclusionLine}}
\vspace{-20pt}
\end{figure}

\textbf{\emph{Experimental Design and Exclusion Line}}. The schematic
figure for the experimental setups are shown in the upper panel of
Fig.~\ref{fig:ExclusionLine}. On the left is interaction region,
where light converts to axions. The light is reflected between two
mirrors to enhance photon number. On the right is detecting crystal,
where the axions convert back to light and then be detected. An opaque
wall blocks the light while axions can cross it freely, as in LSW
experiments. We consider the same laser as reported in ALPS-II\citep{ALPSII},
with effective laser power at $P_{\text{laser}}=150\text{\,kW}$, wavelength
$\lambda=1064\text{\,nm}$ in vacuum, corresponding to photon energy
$\omega=1.17\text{\,eV}$. For 1 year experiment with running time
$3\times10^{7}$s, the total photon number is $N_{\gamma}^{\text{laser}}=2.41\times10^{31}$.
The event number is: 
\begin{align*}
N_{\text{event}}= & N_{\gamma}^{\text{laser}}P_{\text{laser}\rightarrow a}P_{a\rightarrow\gamma}\left(\frac{g_{a\gamma\gamma}}{10^{-7}\text{GeV}^{-1}}\right)^{4}.
\end{align*}
Using $N_{\text{event}}=1$ as criterion, we have $g_{a\gamma\gamma}\ge1.29\times10^{-10}\text{GeV}^{-1}$
for $L=1\text{m}$ and $g_{a\gamma\gamma}\ge1.32\times10^{-11}\text{GeV}^{-1}$
for $L=10\text{m}$, in the limit of $m_{a}\ll\omega$. The complete
exclusion line should be obtained by scanning the $g_{a\gamma\gamma}-m_{a}$
plane, and is shown in the lower panel of Fig.~\ref{fig:ExclusionLine}.

\textbf{\emph{Conclusion}}. In this letter, we present a novel way
to create axions and ALPs in terrestrial experiment. We use optical
laser to interact ionic crystal, and find a coherent enhancement at
the phase-matching condition. 
These coherence is similar to those in previous studies\citep{Buchmuller:1989rb,Henke:1993eda,Yamaji:2017pep,Halliday:2024lca,Matsumoto:2024fzr},
and further increase of the coherence is achieved by the use of thin film layers.
Under our design, the conversion probability
can be 2 orders of magnitude larger than traditional LSW experiment,
as long as the size of the crystal is the same as the magnetic fields
in LSW experiment. We estimate that our proposal is able to push the
axion exclusion line down to $g_{a\gamma\gamma}\gtrsim1.32\times10^{-11}\text{GeV}^{-1}$.

Current coating technique has already enabled the stacking of 1000
layers, with the thickness of single or multilayers from $5\text{\,nm}$
to $10\text{\,\ensuremath{\mu\text{m}}}$, and the size of the film
of tens of square centi-meters\citep{HeliosSputteringTool}. Our design
is therefore possible with current or near-future technique.

\textbf{\emph{Acknowledgments. }}This work is supported by National
Natural Science Foundation of China (No. 12388102), the Strategic
Priority Research Program of the Chinese Academy of Sciences (No.
XDB0890303), the CAS Project for Young Scientists in Basic Research
(No. YSBR060). We thank Yin Hang and Lianghong Yu from SIOM, CAS,
and Zheng Gong from ITP, CAS for helpful discussion.

 \bibliographystyle{apsrev4-2}

%

\onecolumngrid

\appendix
\newpage{}

\section{Axion Production in Coulomb Potentials\label{sec:ClassicalFieldApproach}}

In this appendix, we will consider the interaction between the laser and the ions, 
where the ions are components of some ionic crystal. We mention
that there is a variety of ionic crystals widely used in laser science.
For example, the calcium fluoride ($\text{CaF}_{2}$) crystal is transparent
and can be used as light amplifier. When a laser is injected into
such crystals, its magnetic component will interact with the Coulomb
field of the ions, contributing a non-zero axion source $\boldsymbol{E}\cdot\boldsymbol{B}$.

The Lagrangian for ALP is:

\begin{equation}
\mathcal{L}=\frac{1}{2}\partial_{\mu}a\partial^{\mu}a-\frac{1}{2}m_{a}^{2}a^{2}-\frac{1}{4}F_{\mu\nu}F^{\mu\nu}-\frac{1}{4}g_{a\gamma\gamma}aF_{\mu\nu}\tilde{F}^{\mu\nu},\label{eq:Lagrangian-1}
\end{equation}
where $a$ is the axion field with mass $m_{a}$, $F_{\mu\nu}\equiv\partial_{\mu}A_{\nu}-\partial_{\nu}A_{\mu}$
is the field strength of the electromagnetic field $A_{\mu}$, $\tilde{F}_{\mu\nu}\equiv1/2\varepsilon_{\mu\nu\rho\sigma}F_{\rho\sigma}$
is its dual, and $g_{a\gamma\gamma}$ is the coupling constant with
the dimension of inverse energy. We mention that $-1/4F_{\mu\nu}\tilde{F}^{\mu\nu}=\boldsymbol{E}\cdot\boldsymbol{B}$
where $\boldsymbol{E}$ is the electric field and $\boldsymbol{B}$
is the magnetic field. 
The field equation for $a\left(t,\boldsymbol{r}\right)$ is derived by differentiating the Lagrangian:
\begin{equation}
\left(\partial_{t}^{2}-\nabla^{2}+m_{a}^{2}\right)a=g_{a\gamma\gamma}\boldsymbol{E}\cdot\boldsymbol{B}.\label{eq:axion-EOM:a-gamma-gamma-1}
\end{equation}

Eq.(\ref{eq:axion-EOM:a-gamma-gamma-1}) is an ordinary Klein-Gordon
equation, and the axion field $a\left(t,\boldsymbol{r}\right)$ can
be solved when the external source $\boldsymbol{E}\cdot\boldsymbol{B}$
is known. The axion number produced in classical source is (see, e.g.,
Sec.2.4 of \citep{Peskin:QFT}):

\begin{align}
N_{a}= & \int\frac{\td^{3}\boldsymbol{k}_{a}}{\left(2\pi\right)^{3}}\frac{1}{2E_{a}}\left|\tilde{j}\left(k_{a}^{0},\boldsymbol{k}_{a}\right)\right|_{k_{a}^{0}=E_{a}}^{2},\label{eq:ScalarNumberFromClassicalSource-1}
\end{align}
where $E_{a}=\sqrt{\boldsymbol{k}_{a}^{2}+m_{a}^{2}}$ and $\tilde{j}\left(k_{a}\right)$
is the Fourier transformation of the external source: 
\begin{align*}
\tilde{j}\left(k_{a}^{0},\boldsymbol{k}_{a}\right)= & \int\td t\td^{3}\boldsymbol{r}\e^{\i k_{a}^{0}t}\text{e}^{-\i\boldsymbol{k}_{a}\cdot\boldsymbol{r}}j\left(t,\boldsymbol{r}\right),
\end{align*}
where $j\left(t,\boldsymbol{r}\right)=g_{a\gamma\gamma}\boldsymbol{E}\cdot\boldsymbol{B}$.

\subsection{conversion probability in Medium}

We consider a linearly polarized laser propagating along the $x$-axis.
The magnetic component is then: 
\begin{equation}
\boldsymbol{B}=\left(0,B_{0}\cos\left(\omega t-k_{L}x\right),0\right).
\end{equation}
where $\omega$ is the frequency of the laser, and $\boldsymbol{k}_{L}$
is the corresponding wave vector, $k_{L}\equiv\left|\boldsymbol{k}_{L}\right|$.
In crystals, we have refractive index $n=k_{L}/\omega>1$, i.e. $k_{L}>\omega$.

Assuming that this laser passes through a set of point particles with
charges $\left\{ q_{s}\right\} $ located at $\left\{ \boldsymbol{r}_{s}\right\} $.
The corresponding electric field is:

\begin{equation}
\boldsymbol{E}=\sum_{s}\boldsymbol{E}_{s}=\sum_{s}\frac{q_{s}}{4\pi\left|\boldsymbol{r}-\boldsymbol{r}_{s}\right|^{3}}\left(\boldsymbol{r}-\boldsymbol{r}_{s}\right).
\end{equation}
We mention that $\boldsymbol{r}=\left(x,y,z\right)$. The classical
source for axion production is then $j\left(t,\boldsymbol{r}\right)=\sum_{s}g_{a\gamma\gamma}\boldsymbol{E}_{s}\cdot\boldsymbol{B}$,
and its Fourier transformation is:

\begin{align}
\tilde{j}\left(k_{a}^{0},\boldsymbol{k}_{a}\right)\equiv & \frac{g_{a\gamma\gamma}B_{0}}{4}\delta\left(k_{a}^{0}-\omega\right)\tilde{j}\left(\boldsymbol{k}_{a}\right)\mathcal{T}\left(\boldsymbol{k}_{a};\left\{ \boldsymbol{r}_{s}\right\} \right)
\end{align}
where we have defined:

\begin{align}
\tilde{j}\left(\boldsymbol{k}_{a}\right)=\int\td^{3}\boldsymbol{r}\exp\left[-i\boldsymbol{k}_{a}\cdot\boldsymbol{r}\right]j\left(\boldsymbol{r}\right),\qquad j\left(\boldsymbol{r}\right)= & \frac{\e^{\i k_{L}x}}{\left|\boldsymbol{r}\right|^{3}}y.
\end{align}
and 
\begin{equation}
\mathcal{T}\left(\boldsymbol{k}_{a};\left\{ \boldsymbol{r}_{s}\right\} \right)\equiv\sum_{s}q_{s}\e^{-\i\boldsymbol{k}_{a}\cdot\boldsymbol{r}_{s}}\e^{\i k_{L}x_{s}}.\label{eq:TranslationTerm-1}
\end{equation}

The expression for $\tilde{j}\left(k_{a}^{0},\boldsymbol{k}_{a}\right)$
means that, we can separate the source term into two parts. One is
$\tilde{j}\left(\boldsymbol{k}_{a}\right)$, which contains the information
for one single point charge. Another part, $\mathcal{T}\left(\boldsymbol{k}_{a};\left\{ \boldsymbol{r}_{s}\right\} \right)$,
is a summation of phases, which contains the information for charge
positions. We will in the following refer to $\mathcal{T}$ as ``translation
term''.

The expression for $\tilde{j}\left(\boldsymbol{k}_{a}\right)$ can
be derived by performing a Fourier transformation in cylindrical coordinate(see
\citep{Baddour:2011:2dFourier}), where $k_{ar}=k_{a}\sin\theta$,
$k_{ay}=k_{ar}\cos\phi$ and $k_{az}=k_{ar}\sin\phi$. We have:

\begin{equation}
\tilde{j}\left(\boldsymbol{k}_{a}\right)=-4\pi\i\frac{k_{ay}}{k_{ar}^{2}+\left(k_{ax}-k_{L}\right)^{2}}.
\end{equation}
Substituting into the expression for $\tilde{j}\left(k_{a}^{0},\boldsymbol{k}_{a}\right)$,
and writing in vector form, we have:

\begin{align}
\tilde{j}\left(k_{a}^{0},\boldsymbol{k}_{a}\right)= & -\pi\i g_{a\gamma\gamma}B_{0}\delta\left(k_{a}^{0}-\omega\right)\frac{\Delta k_{y}}{\left(\Delta\boldsymbol{k}^{2}\right)^{2}}\left[\sum_{s}q_{s}\e^{-\i\left(\boldsymbol{k}_{a}-\boldsymbol{k}_{L}\right)\cdot\boldsymbol{r}_{s}}\right]\label{eq:SourceTermFourier:Linear-1}
\end{align}
where $\Delta\boldsymbol{k}=\boldsymbol{k}_{a}-\boldsymbol{k}_{L}$.
Then, using Eq.(\ref{eq:ScalarNumberFromClassicalSource-1}), we can
calculate the axion number. Mention that a special care should be
taken for the delta function. We have:

\begin{equation}
\delta\left(\omega\right)^{2}=\frac{T}{2\pi}\delta\left(\omega\right),
\end{equation}
where $T$ is the large time period during which interaction is present.
We can then integrate over the remaining delta function using $\td^{3}\boldsymbol{k}_{a}=\left|\boldsymbol{k}_{a}\right|^{2}\td\left|\boldsymbol{k}_{a}\right|\sin\theta\td\theta\td\phi$,
and have:

\begin{align}
N_{a}= & \frac{g_{a\gamma\gamma}^{2}B_{0}^{2}}{32\pi^{2}}T\left|\boldsymbol{k}_{a}\right|^{3}\int_{0}^{\pi}\td\theta\int_{0}^{2\pi}\td\phi\frac{\cos^{2}\phi\sin^{3}\theta}{\left(\Delta\boldsymbol{k}^{2}\right)^{2}}\left|\sum_{s}q_{s}\e^{-\i\Delta\boldsymbol{k}\cdot\boldsymbol{r}_{s}}\right|^{2},
\end{align}
where $\left|\boldsymbol{k}_{a}\right|=\sqrt{\omega^{2}-m_{a}^{2}}$.

For photon of energy $\omega$, the number of incoming photon during
time $T$ is:

\begin{equation}
N_{\gamma}=\rho_{\gamma}Sv_{\gamma}T=\frac{B_{0}^{2}}{\omega}S\frac{1}{n}T,
\end{equation}
where $S$ is the laser focal area. The conversion probability is:

\begin{align}
P_{\text{laser}\rightarrow a}\equiv\frac{N_{a}}{N_{\gamma}}= & \frac{g_{a\gamma\gamma}^{2}}{32\pi^{2}S}k_{L}\left|\boldsymbol{k}_{a}\right|^{3}\int_{0}^{\pi}\td\theta\int_{0}^{2\pi}\td\phi\frac{\cos^{2}\phi\sin^{3}\theta}{\left(\Delta\boldsymbol{k}^{2}\right)^{2}}\left|\mathcal{T}\left(\Delta\boldsymbol{k};\left\{ \boldsymbol{r}_{s}\right\} \right)\right|^{2}.\label{eq:TransitionProbability-1}
\end{align}


\subsection{Incident Angle for Coherent Enhancement\label{subsec:Coherent-Enhancement}}

From Eq.(\ref{eq:TransitionProbability-1}), we can see that the conversion
probability is related to $\left|\mathcal{T}\right|^{2}$, where $\mathcal{T}$
is the translation term as defined in Eq.(\ref{eq:TranslationTerm-1}).
This term contains a summation over all charges, 
so it is possible to grow when the crystal volume increases, 
as long as the phase of each charge accumulates rather than cancels out. 
In crystals, the unit cells and the charges distribute regularly, 
so we can rewrite the translation term as:
\begin{align*}
\mathcal{T}= & \sum_{s}q_{s}\e^{-\i\left(\boldsymbol{k}_{a}-\boldsymbol{k}_{L}\right)\cdot\boldsymbol{r}_{s}}\\
= & \left[\sum_{c}q_{c}\e^{-\i\left(\boldsymbol{k}_{a}-\boldsymbol{k}_{L}\right)\cdot\delta\boldsymbol{r}_{c}}\right]\left[\sum_{l}\e^{-\i\left(\boldsymbol{k}_{a}-\boldsymbol{k}_{L}\right)\cdot\boldsymbol{r}_{l}}\right]\\
\equiv & \mathcal{T}_{\text{cell}}\cdot\mathcal{T}_{\text{lattice}}
\end{align*}
where $\boldsymbol{r}_{l}$'s are the coordinates for unit cells,
and $\delta\boldsymbol{r}_{c}$'s are the relative position of particle
$c$ in one cell. The summation of $c$ runs over all particles in
one unit cell, and summation of $l$ runs over all coordinates of
unit cells. We then divide the translation term into the contribution
from cell and from lattice. As we can see, the contribution from cell,
$\mathcal{T}_{\text{cell}}$, is a constant term and do not accumulate
with the increase of crystal size. Therefore, we are more interested
in the $\mathcal{T}_{\text{lattice}}$ term.

In order to coherently enhance the axion production, we want to find
a specific direction, where multiple lattice sites have the same phase
$\i\left(\boldsymbol{k}_{a}-\boldsymbol{k}_{L}\right)\cdot\boldsymbol{r}_{l}$.
To do this, we rotate the crystal, so that the incoming laser is not
perpendicular to the crystal surface, as shown in Fig.\ref{fig:CoherentDirection-DispersionCrystal}.

We consider crystal whose unit cell is cubic. In ``crystal coordinate''
where the cells align along axes, the positions of the unit cells
are:

\begin{equation}
\boldsymbol{r}^{\prime}=\left(i_{1},i_{2},i_{3}\right)d,\qquad i_{1},i_{2},i_{3}\in\mathbb{Z},
\end{equation}
where $d$ is the lattice constant. We use prime notation to denote
coordinate in ``crystal coordinate''. In ``laser coordinate''
where the laser is propagating along $x$-direction, the positions
are:

\begin{equation}
\boldsymbol{r}=\left(i_{1}\cos\alpha+i_{2}\sin\alpha,-i_{1}\sin\alpha+i_{2}\cos\alpha,i_{3}\right)d,
\end{equation}
where $\alpha$ is the inclination angle. We use $\Delta\boldsymbol{k}=\boldsymbol{k}_{a}-\boldsymbol{k}_{L}$
for short notation, and the translation term can be written as:

\begin{align*}
\mathcal{T}_{\text{lattice}}= & \sum_{l}\exp\left[-\i\left(\Delta k_{x}x_{l}+\Delta k_{y}y_{l}+\Delta k_{z}z_{l}\right)\right]\\
= & \sum_{i_{1}=0}^{N_{x}-1}\sum_{i_{2}=0}^{N_{y}-1}\sum_{i_{3}=0}^{N_{z}-1}\exp\left[-\i\Delta k_{x}d\left(i_{1}\cos\alpha+i_{2}\sin\alpha\right)-\i\Delta k_{y}d\left(-i_{1}\sin\alpha+i_{2}\cos\alpha\right)-\i\Delta k_{z}i_{3}d\right]\\
= & \sum_{i_{1}=0}^{N_{x}-1}\exp\left[-\i\left(\Delta k_{x}\cos\alpha-\Delta k_{y}\sin\alpha\right)i_{1}d\right]\sum_{i_{2}=0}^{N_{y}-1}\exp\left[-\i\left(\Delta k_{x}\sin\alpha+\Delta k_{y}\cos\alpha\right)i_{2}d\right]\sum_{i_{3}=0}^{N_{z}-1}\exp\left[-\i\Delta k_{z}i_{3}d\right]
\end{align*}
where $N_{x}$, $N_{y}$ and $N_{z}$ is the number of unit cells
along $x^{\prime}$, $y^{\prime}$ and $z^{\prime}$ direction, respectively.
The summation over all lattice sites can be expressed as three geometric series summations, 
which can be easily done and we have:

\begin{align}
\mathcal{T}_{\text{lattice}}= & \left(\frac{1-\exp\left[-\i\left(\Delta k_{x}\cos\alpha-\Delta k_{y}\sin\alpha\right)N_{x}d\right]}{1-\exp\left[-\i\left(\Delta k_{x}\cos\alpha-\Delta k_{y}\sin\alpha\right)d\right]}\right)\nonumber \\
 & \times\left(\frac{1-\exp\left[-\i\left(\Delta k_{x}\sin\alpha+\Delta k_{y}\cos\alpha\right)N_{y}d\right]}{1-\exp\left[-\i\left(\Delta k_{x}\sin\alpha+\Delta k_{y}\cos\alpha\right)d\right]}\right)\left(\frac{1-\exp\left[-\i\Delta k_{z}N_{z}d\right]}{1-\exp\left[-\i\Delta k_{z}d\right]}\right).
\end{align}
It is apparent that $\mathcal{T}_{\text{lattice}}$ will become large
if one or more denominators becomes zero. Unfortunately, the three
denominators cannot be zero simultaneously. We can choose the first
and the third denominators to be zero, i.e., we want:

\begin{equation}
\Delta k_{x}\cos\alpha-\Delta k_{y}\sin\alpha=0,\qquad k_{z}=0.
\end{equation}

We mention that, when $m_{a}\ll\omega$, we have $\boldsymbol{k}_{a}\approx\omega\left(\cos\theta,\sin\theta\cos\phi,\sin\theta\sin\phi\right)$.
The above relation then becomes:

\begin{align*}
\left(\omega\cos\theta-k_{L}\right)\cos\alpha-\omega\sin\alpha\sin\theta\cos\phi=0, & \qquad\left(a\right)\\
\omega\sin\theta\sin\phi=0. & \qquad\left(b\right)
\end{align*}
From Eq.$\left(b\right)$ we can have $\phi=0$. There is still arbitrary
in determining the incident angle $\alpha$, i.e., as long as we have
determined the incident angle of the laser $\alpha$, we can find
some specific direction$\left(\theta\left(\alpha\right),\phi=0\right)$,
on which $\mathcal{T}_{\text{lattice}}$ reaches a maximum. In this
paper, we will further add a requirement that $\theta=-\alpha$, i.e.,
the axion is propagating along $x^{\prime}$-direction, so that we
have:

\begin{equation}
\cos\alpha=\frac{\omega}{p}=\frac{1}{n}.
\end{equation}
Substituting the value of $\alpha$ and $\theta_{0}=-\alpha$, $\phi_{0}=0$
into the expression for $\mathcal{T}_{\text{lattice}}$, we have:

\begin{equation}
\mathcal{T}_{\text{lattice}}\left(\theta_{0},\phi_{0}\right)=N_{x}\left(\frac{1-\exp\left[\i N_{x}\omega d\tan\alpha\right]}{1-\exp\left[\i\omega d\tan\alpha\right]}\right)N_{z}
\end{equation}
i.e., if we choose the incident angle $\alpha=\arccos\frac{1}{n}$,
there will be a coherence peak for axion outgoing in direction $\left(\theta_{0},\phi_{0}\right)$.
The height of this peak is proportional to $\left|\mathcal{T}_{\text{lattice}}\right|^{2}\propto N_{x}^{2}N_{z}^{2}$,
which means that if the crystal expands on $x^{\prime}$ and $z^{\prime}$
direction, the axion number will increase coherently.

\subsection{Two Plane Waves}

\label{sec:Two-Plane-Waves}

Now, we are going to rotate the coordinate. In the new coordinate,
the cell lattice for the crystal will be aligned along the axis. We
will therefore call the new coordinate ``crystal coordinate''. The
original coordinate will be called ``laser coordinate''. In crystal
coordinate, the notation will be denoted with a prime, $\boldsymbol{r}^{\prime}=\left(x^{\prime},y^{\prime},z^{\prime}\right)$
and $\boldsymbol{p}^{\prime}=\left(p_{x}^{\prime},p_{y}^{\prime},p_{z}^{\prime}\right)$.
The rotation occurs in $x$-$y$ plane and with angle $\delta$. 
The forward and backward transformations are:

\begin{equation}
\left\{ \begin{array}{c}
x^{\prime}=x\cos\delta+y\sin\delta\\
y^{\prime}=-x\sin\delta+y\cos\delta\\
z^{\prime}=z
\end{array}\right.,\qquad\left\{ \begin{array}{c}
x=x^{\prime}\cos\delta-y^{\prime}\sin\delta\\
y=x^{\prime}\sin\delta+y^{\prime}\cos\delta\\
z=z^{\prime}
\end{array}\right..
\end{equation}
Similar transformation applies for all vectors. Substituting into
Eq.(\ref{eq:SourceTermFourier:Linear-1}), we have:

\begin{equation}
\tilde{j}^{\prime}\left(k_{a}^{0},\boldsymbol{k}_{a}^{\prime}\right)=-\pi\i g_{a\gamma\gamma}B_{0}\delta\left(k_{a}^{0}-\omega\right)\frac{\Delta k_{x}^{\prime}\sin\delta+\Delta k_{y}^{\prime}\cos\delta}{\Delta\boldsymbol{k}^{\prime2}}\left[\sum_{s}q_{s}\e^{-\i\Delta\boldsymbol{k}^{\prime}\cdot\boldsymbol{r}_{s}^{\prime}}\right].
\end{equation}

If we consider the reflection of laser on the boundary, as shown in
the lowermost layer in Fig.\ref{fig:CoherentDirection-DispersionCrystal}(c),
there are two plane wave propagating in the crystal, with wavevectors:

\begin{equation}
\boldsymbol{k}_{L1}^{\prime}=n\omega\left(\cos\alpha,\sin\alpha,0\right),\qquad\boldsymbol{k}_{L2}^{\prime}=n\omega\left(\cos\alpha,-\sin\alpha,0\right).
\end{equation}
The corresponding rotation angles are $\delta_{1}=\alpha$ and $\delta_{2}=-\alpha$,
and we have:

\begin{align}
\tilde{j}_{1}^{\prime}\left(k_{a}^{0},\boldsymbol{k}_{a}^{\prime}\right)= & -\pi\i g_{a\gamma\gamma}B_{0}\delta\left(k_{a}^{0}-\omega\right)\frac{\Delta k_{1x}^{\prime}\sin\alpha+\Delta k_{1y}^{\prime}\cos\alpha}{\Delta\boldsymbol{k}_{1}^{\prime2}}\left[\sum_{s}q_{s}\e^{-\i\Delta\boldsymbol{k}_{1}^{\prime}\cdot\boldsymbol{r}_{s}^{\prime}}\right]\\
\tilde{j}_{2}^{\prime}\left(k_{a}^{0},\boldsymbol{k}_{a}^{\prime}\right)= & -\pi\i g_{a\gamma\gamma}B_{0}\delta\left(k_{a}^{0}-\omega\right)\frac{-\Delta k_{2x}^{\prime}\sin\alpha+\Delta k_{2y}^{\prime}\cos\alpha}{\Delta\boldsymbol{k}_{2}^{\prime2}}\left[\sum_{s}q_{s}\e^{-\i\Delta\boldsymbol{k}_{2}^{\prime}\cdot\boldsymbol{r}_{s}^{\prime}}\right]
\end{align}
where we have defined $\Delta\boldsymbol{k}_{1}^{\prime}=\boldsymbol{k}_{a}^{\prime}-\boldsymbol{k}_{L1}^{\prime}$,
$\Delta\boldsymbol{k}_{2}^{\prime}=\boldsymbol{k}_{a}^{\prime}-\boldsymbol{k}_{L2}^{\prime}$.

Substituting into Eq.(\ref{eq:ScalarNumberFromClassicalSource-1}),
the axion number is (we omit the prime notation from now on for short
notation, and the following expression is in crystal coordinate):

\begin{align}
N_{a}= & \int\frac{\td^{3}\boldsymbol{q}}{\left(2\pi\right)^{3}}\frac{1}{2E_{a}}\left|\tilde{j}_{1}\left(k_{a}^{0},\boldsymbol{k}_{a}\right)+\tilde{j}_{2}\left(k_{a}^{0},\boldsymbol{k}_{a}\right)\right|_{k_{a}^{0}=E_{a}}^{2}\nonumber \\
= & \pi^{2}g_{a\gamma\gamma}^{2}B_{0}^{2}\frac{T}{2\left(2\pi\right)^{4}}\sqrt{\omega^{2}-m_{a}^{2}}\int_{0}^{\pi}\td\theta\int_{0}^{2\pi}\td\phi\nonumber \\
 & \times\sin\theta\left|\frac{\Delta k_{1x}\sin\alpha+\Delta k_{1y}\cos\alpha}{\Delta\boldsymbol{k}_{1}^{2}}\sum_{s}q_{s}\e^{-\i\Delta\boldsymbol{k}_{1}\cdot\boldsymbol{r}_{s}}+\frac{-\Delta k_{2x}\sin\alpha+\Delta k_{2y}\cos\alpha}{\Delta\boldsymbol{k}_{2}^{2}}\sum_{s}q_{s}\e^{-\i\Delta\boldsymbol{k}_{2}\cdot\boldsymbol{r}_{s}}\right|^{2}
\end{align}

The laser comes from the leftmost boundary, and the photon number
in corresponding time period is:

\begin{equation}
N_{\gamma}=\rho_{\gamma}Sv_{\gamma}T=\frac{B_{0}^{2}}{\omega}S\frac{1}{n}T
\end{equation}
So we have the transition rate:

\begin{align}
P_{\text{laser}\rightarrow a}= & \frac{g_{a\gamma\gamma}^{2}}{32\pi^{2}}\frac{n\omega\left|\boldsymbol{k}_{a}\right|}{S}\int_{0}^{\pi}\td\theta\int_{0}^{2\pi}\td\phi\nonumber \\
 & \times\sin\theta\left|\frac{\Delta k_{1x}\sin\alpha+\Delta k_{1y}\cos\alpha}{\Delta\boldsymbol{k}_{1}^{2}}\sum_{s}q_{s}\e^{-\i\Delta\boldsymbol{k}_{1}\cdot\boldsymbol{r}_{s}}+\frac{-\Delta k_{2x}\sin\alpha+\Delta k_{2y}\cos\alpha}{\Delta\boldsymbol{k}_{2}^{2}}\sum_{s}q_{s}\e^{-\i\Delta\boldsymbol{k}_{2}\cdot\boldsymbol{r}_{s}}\right|^{2}\label{eqn:TwoWaveProbability}
\end{align}

\section{Reconversion in Waveguide\label{sec:Reconversion-in-Waveguide}}

We have derived the transition probability from laser to axion. In
order to detect those axion, we have to study its conversion back
to light signal. With the introducing of axion, the Maxwell equations
for the electro-magnetic field are(see, e.g., Ref.\citep{An:2024kns}):
\begin{align*}
\nabla\cdot\boldsymbol{E}= & \frac{\rho}{\varepsilon}-cg_{a\gamma\gamma}\boldsymbol{B}\cdot\nabla a\\
\nabla\cdot\boldsymbol{B}= & 0\\
\nabla\times\boldsymbol{E}= & -\frac{\partial}{\partial t}\boldsymbol{B}\\
\nabla\times\boldsymbol{B}= & \mu\varepsilon\frac{\partial}{\partial t}\boldsymbol{E}+\mu\boldsymbol{J}+\frac{g_{a\gamma\gamma}}{c}\left[\frac{\partial a}{\partial t}\boldsymbol{B}-\boldsymbol{E}\times\nabla a\right].
\end{align*}
The electro-magnetice fields inside the crystal can be splitted into
two parts: the static, external Coulomb fields inside the crystal,
and the propagating fields converted from axion:

\[
\boldsymbol{E}=\boldsymbol{E}^{\text{ext}}+\boldsymbol{E}^{\text{prop}},\qquad\boldsymbol{B}=\boldsymbol{B}^{\text{prop}},
\]
where the $\boldsymbol{E}^{\text{ext}}$ is provided by the ions in
ionic crystals. When considering the reconversion process, we neglect
the possibility of light transits to axion again. Also, since the
regenerated light is very weak and $g_{a\gamma\gamma}$ is very small,
we can neglect the $g_{a\gamma\gamma}\boldsymbol{B}^{\text{prop}}$
and $g_{a\gamma\gamma}\boldsymbol{E}^{\text{prop}}$ terms. The current
$\boldsymbol{J}$ inside crystal is also ignored. The equations then
become (we suppress the upperscript ``prop'' for shorter notation):

\begin{align*}
\nabla\times\boldsymbol{E}= & -\frac{\partial\boldsymbol{B}}{\partial t}\\
\nabla\times\boldsymbol{B}= & \mu\varepsilon\frac{\partial\boldsymbol{E}}{\partial t}-g_{a\gamma\gamma}\boldsymbol{E}^{\text{ext}}\times\nabla a\\
\nabla\cdot\boldsymbol{E}= & 0\\
\nabla\cdot\boldsymbol{B}= & 0
\end{align*}
As can be seen, the contribution from axion can be regarded as an
effecitive current $\boldsymbol{J}_{\text{eff}}$, defined as
\begin{equation}
\boldsymbol{J}_{\text{eff}}\left(t,\boldsymbol{r}\right)=-\frac{1}{\mu}g_{a\gamma\gamma}\boldsymbol{E}^{\text{ext}}\times\nabla a.\label{eq:axion-effective-current}
\end{equation}

The regenerated light will be reflected by the boundary of the crystal,
which can be regarded as a waveguide. We consider a rectangular waveguide
propagating along $x$-axis, and the cross-section area is $y\in\left[0,L_{y}\right]$,
$z\in\left[0,L_{z}\right]$. The phase-match condition from axion
production region indicate that the regenerated light will excite
the transverse electric(TE) mode of the waveguide, so that the longitudinal
electric field, $E_{x}$ vanishes. The expansion modal of the modes
are\citep{Jackson1999:ClassicalElectrodynamics}:

\begin{align}
E_{ymn}\left(y,z\right)= & -\frac{2\pi n}{\gamma_{mn}L_{z}\sqrt{L_{y}L_{z}}}\cos\left(\frac{m\pi y}{L_{y}}\right)\sin\left(\frac{n\pi z}{L_{z}}\right),\label{eq:Jackson:1999EM:8.136a-1}\\
E_{zmn}\left(y,z\right)= & \frac{2\pi m}{\gamma_{mn}L_{y}\sqrt{L_{y}L_{z}}}\sin\left(\frac{m\pi y}{L_{y}}\right)\cos\left(\frac{n\pi z}{L_{z}}\right),\label{eq:Jackson:1999EM:8.136b-1}\\
H_{xmn}\left(y,z\right)= & -\frac{2\i\gamma_{mn}}{k_{mn}Z_{mn}\sqrt{L_{y}L_{z}}}\cos\left(\frac{m\pi y}{L_{y}}\right)\cos\left(\frac{n\pi z}{L_{z}}\right),\label{eq:Jackson:1999EM:8.136c-1}
\end{align}
where $\left(m,n\right)$ are two indices indicating the different
mode, $\gamma_{mn}$, $k_{mn}$ and $Z_{mn}$ are given by:

\begin{align*}
\gamma_{mn}^{2}= & \pi^{2}\left(\frac{m^{2}}{L_{y}^{2}}+\frac{n^{2}}{L_{z}^{2}}\right),\\
k_{mn}^{2}= & \mu\varepsilon\omega^{2}-\gamma_{mn}^{2},\\
Z_{mn}= & \mu\omega/k_{mn}.
\end{align*}

Since $L_{y}$ and $L_{y}$ are large, the integers $m$ and $n$
can be regarded as continuous variable, and so are $\gamma_{mn}$
and $k_{mn}$. As we discussed in the main text, the coherence appears
for $n=0$ and $k_{m0}=\omega$. The electric and magnetic component,
Eq.(\ref{eq:Jackson:1999EM:8.136a-1}-\ref{eq:Jackson:1999EM:8.136c-1})
should then be divided by $\sqrt{2}$ to satisfy the normalization
condition. Mention that the $E_{y}$ component disappears for $n=0$
mode. 

The electric field for $\left(m,0\right)$ mode is then:

\begin{align}
\boldsymbol{E}\left(x,y,z,t\right)= & A_{m0}E_{zm0}\left(y\right)\hat{\boldsymbol{z}}\e^{-\i\omega t+\i k_{m0}x}
\end{align}
where the coefficient $A_{m0}$ can be calculated as:
\begin{align}
A_{m0}= & -\frac{Z_{m0}}{2}\int_{V}\td^{3}\boldsymbol{r}\boldsymbol{J}_{\text{eff}}\left(\boldsymbol{r}\right)\cdot\hat{\boldsymbol{z}}E_{zm0}\left(y\right)\e^{-\i k_{m0}x},\label{eq:regenerate:waveguide-coeff}
\end{align}
where $V$ is a large volume containing all the sources $\boldsymbol{J}_{\text{eff}}$.
Notice the minus sign of $\e^{-\i k_{m0}x}$ in Eq.(\ref{eq:regenerate:waveguide-coeff}).
The time component of $\boldsymbol{J}_{\text{eff}}$ have been extracted,
$\boldsymbol{J}_{\text{eff}}\left(t,\boldsymbol{r}\right)=\boldsymbol{J}_{\text{eff}}\left(\boldsymbol{r}\right)\e^{-\i\omega t}.$

Now we are going to study the source term as in Eq.(\ref{eq:axion-effective-current}).
The external electric field is provided by the ions inside the ionic
crystal. For point particles in free space, the electric field is:
\begin{equation}
\boldsymbol{E}^{\text{ext}}\left(\boldsymbol{r}\right)=\sum_{s}\frac{Q_{s}}{4\pi\varepsilon_{0}\left|\boldsymbol{r}-\boldsymbol{r}_{s}\right|^{3}}\left(\boldsymbol{r}-\boldsymbol{r}_{s}\right).\label{eq:E-field:freespace}
\end{equation}
Inside medium and waveguide, the electric field should in principle
be modified by the medium and also by the boundary condition of the
waveguide. However, since the crystal is charge neutral, the electric
field far away from any charge will be shielded by another nearby
charge with opposite sign. The interaction is significant only in
the neighbouhood of the charges, where the boundary condition and
medium effect is not important. We will therefore use the free-space
form of the charges. 

For the axion field, we consider it to be a plane wave propagating
along $x$-axis:

\begin{equation}
a\left(t,\boldsymbol{r}\right)=a_{0}\e^{-\i\omega t+\i\boldsymbol{p}_{a}\cdot\boldsymbol{r}}
\end{equation}
where $\boldsymbol{p}_{a}=\left(p_{a},0,0\right)$. Substituting into
Eq.(\ref{eq:axion-effective-current}), we have the source term: 

\begin{equation}
\boldsymbol{J}_{\text{eff}}\left(\boldsymbol{r}\right)=-\frac{\i}{\mu}g_{a\gamma\gamma}a_{0}\e^{\i\boldsymbol{p}_{a}\cdot\boldsymbol{r}}\boldsymbol{E}^{\text{ext}}\left(\boldsymbol{r}\right)\times\boldsymbol{p}_{a}.
\end{equation}

Then, using Eq.(\ref{eq:regenerate:waveguide-coeff}), we can derive
the amplitude:

\begin{align}
A_{m0}= & \frac{\sqrt{2}}{2}\frac{\i g_{a\gamma\gamma}a_{0}}{\sqrt{L_{y}L_{z}}}\frac{\omega}{k_{m0}\varepsilon_{0}}p_{a}\sum_{s}Q_{s}\e^{\i\left(p_{a}-\omega\right)x_{s}+\i q_{y}y_{s}}\frac{q_{y}}{\left(p_{a}-\omega\right)^{2}+q_{y}^{2}}
\end{align}
where $q_{y}=m\pi/L_{y}$. By carefully choosing the mode, we can
have $k_{m0}=\omega$, and $q_{y}=\sqrt{n^{2}-1}\omega$, to satisfy
the coherence.

The transverse magnetic field is:

\begin{equation}
\boldsymbol{H}_{t}=\frac{1}{Z}\hat{\boldsymbol{x}}\times\boldsymbol{E}_{t}.
\end{equation}
For $\left(m,0\right)$ mode, the electric field only has $z$-component,
so the transverse magnetic field only has $y$-component (together
with longitudinal $H_{x}$ component). The Poynting vector of the
regenerated light is then (an extra $1/2$ factor for averaging over
time):

\begin{align}
\left\langle \boldsymbol{S}\right\rangle = & \frac{1}{2}\boldsymbol{E}\times\boldsymbol{H}^{\ast}=\frac{1}{2Z_{m0}}E_{z}^{2}\hat{\boldsymbol{x}}.
\end{align}
For a cross sectional area $S=L_{y}L_{z}$, we integrate over $S$
and obtain the averaged power:
\begin{align}
\left\langle P_{\text{waveguide}}\right\rangle = & \int\td y\td z\left\langle S_{x}\right\rangle 
\end{align}

For a plane wave of axion, the time-averaged energy density is:

\begin{equation}
\left\langle \rho_{a}\right\rangle =\frac{1}{2}a_{0}^{2}\left(\boldsymbol{p}_{a}^{2}+m_{a}^{2}\right)=\frac{1}{2}a_{0}^{2}\omega^{2},
\end{equation}
and the power for axion across cross sectional area $S$ is:
\begin{equation}
\left\langle P_{a}\right\rangle =\frac{1}{2}a_{0}^{2}\omega^{2}v_{a}S
\end{equation}
So the conversion probability is:

\begin{equation}
P_{a\rightarrow\gamma}\equiv\frac{\left\langle P_{\text{waveguide}}\right\rangle }{\left\langle P_{a}\right\rangle }=\frac{1}{2\mu}\frac{g_{a\gamma\gamma}^{2}}{\varepsilon_{0}^{2}\omega^{2}v_{a}S^{2}}\frac{p_{a}^{2}q_{y}^{2}}{\boldsymbol{k}^{4}}\left|\sum_{s}Q_{s}\e^{\i\left(p_{a}-\omega\right)x_{s}+\i q_{y}y_{s}}\right|^{2}\label{eq:P_re:waveguide}
\end{equation}
where for maximum coherence, we have already require:

\begin{align*}
k_{m0}= & \omega,\\
Z_{m0}= & \mu.
\end{align*}

\section{Contribution from In-Atom Electric Field\label{sec:In-Atom}}

Throughout this paper, we have been treating the ions as point charges.
However, the ions have finite size and the electric field inside those
ions can be much larger than the inverse-square Coulomb field, since
the nucleous have larger charge. One may expect that these larger
electric field might contribute to extra axion production. However,
in this section, we will show that those contribution is negligible
for long-wavelength optical light.

For an ion whose nuclear charge is $Z_{s}$, total charge number is
$Q_{s}$ (i.e., with $Z_{s}-Q_{s}$ electrons) and radius is $R_{s}$,
we model it as a pure nucleous surrounded by electron cloud. The electric
field from the nucleous is:

\[
\boldsymbol{E}_{s}^{\text{nucl}}=\frac{Z_{s}e}{4\pi\left|\boldsymbol{r}-\boldsymbol{r}_{s}\right|^{3}}\left(\boldsymbol{r}-\boldsymbol{r}_{s}\right).
\]
The electron cloud is considered to be uniformly distributed inside
a sphere with radius $R_{s}$. Inside the sphere, the electric field
can be derived using Gauss law, while the electric field outside the
sphere is the same as point charge: 
\begin{align*}
\boldsymbol{E}_{s}^{\text{e}}= & -\frac{\left(Z_{s}-Q_{s}\right)e}{4\pi R_{s}^{3}}\left(\boldsymbol{r}-\boldsymbol{r}_{s}\right)\Theta\left(R_{s}-\left|\boldsymbol{r}-\boldsymbol{r}_{s}\right|\right)-\frac{\left(Z_{s}-Q_{s}\right)e}{4\pi\left|\boldsymbol{r}-\boldsymbol{r}_{s}\right|^{3}}\left(\boldsymbol{r}-\boldsymbol{r}_{s}\right)\Theta\left(\left|\boldsymbol{r}-\boldsymbol{r}_{s}\right|-R_{s}\right)\\
= & -\frac{\left(Z_{s}-Q_{s}\right)e}{4\pi}\left(\boldsymbol{r}-\boldsymbol{r}_{s}\right)\Theta\left(R_{s}-\left|\boldsymbol{r}-\boldsymbol{r}_{s}\right|\right)\left[\frac{1}{R_{s}^{3}}-\frac{1}{\left|\boldsymbol{r}-\boldsymbol{r}_{s}\right|^{3}}\right]-\frac{\left(Z_{s}-Q_{s}\right)e}{4\pi\left|\boldsymbol{r}-\boldsymbol{r}_{s}\right|^{3}}\left(\boldsymbol{r}-\boldsymbol{r}_{s}\right),
\end{align*}
where $\Theta$ is the step function.

The total electric field is then the sum of contribution from nucleous
and electron:

\begin{align*}
\boldsymbol{E}_{s}= & \boldsymbol{E}_{s}^{\text{nucl}}+\boldsymbol{E}_{s}^{\text{e}}\\
= & \frac{Q_{s}e}{4\pi\left|\boldsymbol{r}-\boldsymbol{r}_{s}\right|^{3}}\left(\boldsymbol{r}-\boldsymbol{r}_{s}\right)-\frac{\left(Z_{s}-Q_{s}\right)e}{4\pi}\left(\boldsymbol{r}-\boldsymbol{r}_{s}\right)\Theta\left(R_{s}-\left|\boldsymbol{r}-\boldsymbol{r}_{s}\right|\right)\left[\frac{1}{R_{s}^{3}}-\frac{1}{\left|\boldsymbol{r}-\boldsymbol{r}_{s}\right|^{3}}\right]\\
\equiv & \boldsymbol{E}_{s}^{\text{point}}+\boldsymbol{E}_{s}^{\text{in-ion}},
\end{align*}
where we have defined: 
\begin{align*}
\boldsymbol{E}_{s}^{\text{point}}= & \frac{Q_{s}e}{4\pi\left|\boldsymbol{r}-\boldsymbol{r}_{s}\right|^{3}}\left(\boldsymbol{r}-\boldsymbol{r}_{s}\right),\\
\boldsymbol{E}_{s}^{\text{in-ion}}= & -\frac{\left(Z_{s}-Q_{s}\right)e}{4\pi}\left(\boldsymbol{r}-\boldsymbol{r}_{s}\right)\Theta\left(R_{s}-\left|\boldsymbol{r}-\boldsymbol{r}_{s}\right|\right)\left[\frac{1}{R_{s}^{3}}-\frac{1}{\left|\boldsymbol{r}-\boldsymbol{r}_{s}\right|^{3}}\right].
\end{align*}
It is clear that $\boldsymbol{E}_{s}^{\text{point}}$ is the point
charge contribution as we have used in the main text and in Appendix.\ref{sec:ClassicalFieldApproach},
and $\boldsymbol{E}_{s}^{\text{in-ion}}$ is the modification from
the electric field inside ions.

For simplicity, we consider a circularly polarized laser propagating
along $x$-direction inside the medium. The corresponding $\boldsymbol{B}$
component is: 
\[
\boldsymbol{B}=\left(0,B_{0}\cos\left(\omega t-k_{L}x\right),B_{0}\sin\left(\omega t-k_{L}x\right)\right),
\]
where $k_{L}=n\omega$ with refracive index $n>1$.

The classical source for axion in Eq.(\ref{eq:ScalarNumberFromClassicalSource})
is then

\begin{align*}
j\left(t,\boldsymbol{r}\right)= & \sum_{s}g_{a\gamma\gamma}\boldsymbol{E}_{s}^{\text{point}}\cdot\boldsymbol{B}+\sum_{s}g_{a\gamma\gamma}\boldsymbol{E}_{s}^{\text{in-ion}}\cdot\boldsymbol{B}\\
\equiv & j^{\text{point}}\left(t,\boldsymbol{r}\right)+j^{\text{in-ion}}\left(t,\boldsymbol{r}\right),
\end{align*}
where the $j^{\text{point}}\left(t,\boldsymbol{r}\right)$ term can
be derived following Appendix.\ref{sec:ClassicalFieldApproach}, except
that we use circularly polarized laser here. The Fourier transformation
of $j^{\text{point}}$ should then be:

\begin{align}
\tilde{j}^{\text{point}}\left(k_{a}^{0},\boldsymbol{k}_{a}\right)= & -\frac{\pi}{2}g_{a\gamma\gamma}B_{0}e\i\e^{\i\phi}\delta\left(k_{a}^{0}-\omega\right)\sum_{s}Q_{s}\e^{-\i\left(\boldsymbol{k}_{a}-\boldsymbol{k}_{L}\right)\cdot\boldsymbol{r}_{s}}\frac{2\left|\boldsymbol{k}_{a}\right|\sin\theta}{\left(\boldsymbol{k}_{a}-\boldsymbol{k}_{L}\right)^{2}},\label{eq:jtildep:circular}
\end{align}
where $\theta$ is the polar angle between $\boldsymbol{k}_{a}$ and
$x$-axis (mention that $\boldsymbol{k}_{L}$ is along $x$-axis),
and $\phi$ is the azimuthal angle. The angles are defined through
$k_{ax}=\left|\boldsymbol{k}_{a}\right|\cos\theta$, $k_{ay}=\left|\boldsymbol{k}_{a}\right|\sin\theta\cos\phi$,
$k_{az}=\left|\boldsymbol{k}_{a}\right|\sin\theta\sin\phi$.

In order to calculate the axion conversion probability, we need then
derive $j^{\text{in-ion}}$ term. The Fourier transformation of the
source is:

\begin{align*}
\tilde{j}^{\text{in-ion}}\left(k_{a}^{0},\boldsymbol{k}_{a}\right)= & -\int\td t\td^{3}\boldsymbol{r}\exp\left[\i tk_{a}^{0}-\i\boldsymbol{k}_{a}\cdot\boldsymbol{r}\right]j^{\text{in-ion}}\left(t,\boldsymbol{r}\right)\\
= & -\frac{g_{a\gamma\gamma}B_{0}e}{4}\delta\left(k_{a}^{0}-\omega\right)\sum_{s}\left(Z_{s}-Q_{s}\right)\e^{-\i\boldsymbol{k}_{a}\cdot\boldsymbol{r}_{s}}\e^{\i k_{L}x_{s}}\\
 & \times\left\{ \int\td^{3}\boldsymbol{r}\e^{-\i\boldsymbol{k}_{a}\cdot\boldsymbol{r}}\e^{\i k_{L}x}\left(y+\i z\right)\left(\frac{1}{R_{s}^{3}}-\frac{1}{\left|\boldsymbol{r}\right|^{3}}\right)\Theta\left(R_{s}-\left|\boldsymbol{r}\right|\right)\right\} 
\end{align*}

The terms in the curly bracket is the contribution for a single ion
at origin. The integration over $\td^{3}\boldsymbol{r}$ is the corresponding
Fourier transformation. For short notation, we define:

\begin{align}
j_{s}^{\text{in-ion}}\left(\boldsymbol{r}\right)= & \e^{\i k_{L}x}\left(y+\i z\right)\left(\frac{1}{R_{s}^{3}}-\frac{1}{\left|\boldsymbol{r}\right|^{3}}\right)\Theta\left(R_{s}-\left|\boldsymbol{r}\right|\right),\nonumber \\
\tilde{j}_{s}^{\text{in-ion}}\left(\boldsymbol{k}_{a}\right)= & \int\td^{3}\boldsymbol{r}\exp\left[-\i\boldsymbol{k}_{a}\cdot\boldsymbol{r}\right]j_{s}^{\text{in-ion}}\left(\boldsymbol{r}\right),\label{eq:jsk}
\end{align}
and we have:

\begin{align}
\Rightarrow\tilde{j}^{\text{in-ion}}\left(k_{a}^{0},\boldsymbol{k}_{a}\right)\equiv & -\frac{g_{a\gamma\gamma}B_{0}e}{4}\delta\left(k_{a}^{0}-\omega\right)\sum_{s}\left(Z_{s}-Q_{s}\right)\e^{-\i\boldsymbol{k}_{a}\cdot\boldsymbol{r}_{s}}\e^{\i k_{L}x_{s}}\tilde{j}_{s}^{\text{in-ion}}\left(\boldsymbol{k}_{a}\right)\label{eq:jtildep:LaserAtom-2-1-1}
\end{align}

We are going to use cylindrical coordinate, $\boldsymbol{r}=\left(x,y,z\right)=\left(x,\rho,\psi\right)$,
where $\rho=\sqrt{y^{2}+z^{2}}$, $y=\rho\cos\psi$ and $z=\rho\sin\psi$.
The Fourier transformation can also be written in cylindrical coordinate,
i.e., $\boldsymbol{k}_{a}=\left(k_{ax},k_{ay},k_{az}\right)=\left(k_{ax},k_{a\rho},\phi\right)$,
where $k_{a\rho}=\sqrt{k_{ay}^{2}+k_{az}^{2}}$, $k_{ay}=k_{a\rho}\cos\phi$
and $k_{az}=k_{a\rho}\sin\phi$. 
As shown in Ref.\citep{Baddour:2011:2dFourier}, the Fourier transformation in cylindrical coordinate should be written
as:

\begin{align}
\tilde{j}_{s}^{\text{in-ion}}\left(k_{ax},k_{a\rho},\phi\right)= & -2\pi\i\e^{\i\phi}\int\td x\e^{-\i\left(k_{ax}-k_{L}\right)x}\int_{0}^{\infty}\rho^{2}\td\rho\left(\frac{1}{R_{s}^{3}}-\frac{1}{\left(x^{2}+\rho^{2}\right)^{3/2}}\right)\Theta\left(R_{s}-\left(x^{2}+\rho^{2}\right)^{1/2}\right)J_{1}\left(k_{a\rho}\rho\right),\label{eq:2dpolarFourier}
\end{align}
where $J_{1}$ is the first order Bessel function.

This integration cannot be integrated analytically. However, if we
use optical light, the wavelength $\lambda$ is far larger than the
atom radius $R_{s}$. We therefore have $\left|\boldsymbol{k}_{a}\right|R_{s},k_{ax}R_{s},k_{a\rho}R_{s}\ll1$
and the Bessel function approaches its asymptotic expression:

\[
J_{\alpha}\left(z\right)\sim\frac{1}{\Gamma\left(\alpha+1\right)}\left(\frac{z}{2}\right)^{\alpha},\qquad0<z\ll\sqrt{\alpha+1}.
\]
We can then perform the integration in Eq.(\ref{eq:2dpolarFourier})
and approximately have:

\begin{align*}
\tilde{j}_{s}^{\text{in-ion}}\left(k_{ax},k_{a\rho},\phi\right)\approx & -4\pi\i\e^{\i\phi}\frac{k_{\rho}}{\left(k_{ax}-k_{L}\right)^{5}R_{s}^{3}}\left[\left(k_{ax}-k_{L}\right)^{3}R_{s}^{3}+3\left(k_{ax}-k_{L}\right)R_{s}\cos\left(\left(k_{ax}-k_{L}\right)R_{s}\right)-3\sin\left(\left(k_{ax}-k_{L}\right)R_{s}\right)\right]
\end{align*}

As we have stated, $\left(k_{ax}-k_{L}\right)R_{s}\ll1$. We can then
further approximate the $\sin$ and $\cos$ function in this limit
and have:

\begin{align}
\tilde{j}_{s}^{\text{in-ion}}\left(k_{ax},k_{a\rho},\phi\right)\approx & -4\pi\i\e^{\i\phi}k_{\rho}R_{s}^{2}\label{eq:j_in-ion}
\end{align}

Substituting Eq.(\ref{eq:j_in-ion}) into Eq.(\ref{eq:jtildep:LaserAtom-2-1-1}),
the compelete Fourier transformation for in-ion contribution is:

\begin{align*}
\tilde{j}^{\text{in-ion}}\left(k_{a}^{0},\boldsymbol{k}_{a}\right)= & \pi\i g_{a\gamma\gamma}B_{0}e\e^{\i\phi}\delta\left(k_{a}^{0}-\omega\right)\sum_{s}\left(Z_{s}-Q_{s}\right)\e^{-\i\left(\boldsymbol{k}_{a}-\boldsymbol{k}_{L}\right)\cdot\boldsymbol{r}_{s}}\left|\boldsymbol{k}_{a}\right|\sin\theta R_{s}^{2}
\end{align*}
The point-charge contribution is given in Eq.(\ref{eq:jtildep:circular}),
and the total source term is 
\begin{align}
\tilde{j}\left(k_{a}^{0},\boldsymbol{k}_{a}\right)= & \tilde{j}^{\text{point}}\left(k_{a}^{0},\boldsymbol{k}_{a}\right)+\tilde{j}^{\text{in-ion}}\left(k_{a}^{0},\boldsymbol{k}_{a}\right)\nonumber \\
= & -\pi g_{a\gamma\gamma}B_{0}e\i\e^{\i\phi}\delta\left(k_{a}^{0}-\omega\right)\frac{\left|\boldsymbol{k}_{a}\right|\sin\theta}{\left(\Delta\boldsymbol{k}\right)^{2}}\sum_{s}\e^{-\i\Delta\boldsymbol{k}\cdot\boldsymbol{r}_{s}}\left\{ Q_{s}-\left(Z_{s}-Q_{s}\right)\left(\Delta\boldsymbol{k}\right)^{2}R_{s}^{2}\right\} ,\label{eqn:j-tilde}
\end{align}
where $\Delta\boldsymbol{k}=\boldsymbol{k}_{a}-\boldsymbol{k}_{L}$
is the momentum transfer. The first term in the curly bracket is the
result from point-charge approximation, and the second term in the
curly bracket comes from the modification of in-ion electric field.
Mention that the wavelength of the laser is of order $\lambda\approx2\pi/|\Delta\boldsymbol{k}|\sim10^{-6}\text{m}$,
while the radius of atom is of order $R_{s}\sim10^{-10}-10^{-9}\text{m}$.
The modification from the in-ion electric field is of order $\left(R_{s}/\lambda\right)^{2}$.

However, considering that the crystal is charge neutral, there is
a cancellation in the point-charge term. As in Eq.(\ref{eq:T_cell_lat})
in the main text, the point-charge term can be split into two component:
\[
\sum_{s}\e^{-\i\Delta\boldsymbol{k}_{a}\cdot\boldsymbol{r}_{s}}Q_{s}=\left\{ \sum_{c}\e^{-\i\Delta\boldsymbol{k}\cdot\delta\boldsymbol{r}_{c}}Q_{c}\right\} \left\{ \sum_{l}\e^{-\i\Delta\boldsymbol{k}\cdot\boldsymbol{r}_{l}}\right\} ,
\]
where the first term is the contribution from a single cell and the
subscript $c$ runs over all charges inside one cell, the second term
is the contribution from entire lattice and the subscript $l$ runs
over all the lattice cells. $\delta\boldsymbol{r}_{c}$ is the relative
coordinate of charges inside one cell. Apparently, since the crystal
is charge neutral, we have $\sum_{c}Q_{c}=0$. However, the spatial
translation of charges provide extra phases, and we have: 
\[
\sum_{c}\e^{-\i\Delta\boldsymbol{k}\cdot\delta\boldsymbol{r}_{c}}Q_{c}\approx\i\left|\Delta\boldsymbol{k}\right|d,
\]
where $d$ is the crystal lattice constant and we have used the expansion
$\exp\left(-\i\Delta\boldsymbol{k}\cdot\delta\boldsymbol{r}_{c}\right)\approx1-\left(\i\Delta\boldsymbol{k}\cdot\delta\boldsymbol{r}_{c}\right)$.
Considering that we have $|\Delta\boldsymbol{k}|\approx2\pi/\lambda$,
the first term in the curly bracket of (\ref{eqn:j-tilde}) damps
as $d/\lambda$. The ion radius $R_{s}$ is comparable to lattice
constant $d$, so the second term in the curly bracket of (\ref{eqn:j-tilde})
is smaller than the first term by a factor of $d/\lambda\approx10^{-3}$,
so we can safely omit the second term, i.e. the contribution from
in-ion electric field can be ignored.

We also mention that, if we take $Q_{s}=0$, the result corresponds
to axion production from charge-neutral atoms. Therefore, the contribution
from the supporting material in Fig.\ref{fig:CoherentDirection-DispersionCrystal}(c)
is negligible due to the $R_{s}/\lambda$ suppression. 

\begin{thebibliography}{30}%
\makeatletter
\providecommand \@ifxundefined [1]{%
 \@ifx{#1\undefined}
}%
\providecommand \@ifnum [1]{%
 \ifnum #1\expandafter \@firstoftwo
 \else \expandafter \@secondoftwo
 \fi
}%
\providecommand \@ifx [1]{%
 \ifx #1\expandafter \@firstoftwo
 \else \expandafter \@secondoftwo
 \fi
}%
\providecommand \natexlab [1]{#1}%
\providecommand \enquote  [1]{``#1''}%
\providecommand \bibnamefont  [1]{#1}%
\providecommand \bibfnamefont [1]{#1}%
\providecommand \citenamefont [1]{#1}%
\providecommand \href@noop [0]{\@secondoftwo}%
\providecommand \href [0]{\begingroup \@sanitize@url \@href}%
\providecommand \@href[1]{\@@startlink{#1}\@@href}%
\providecommand \@@href[1]{\endgroup#1\@@endlink}%
\providecommand \@sanitize@url [0]{\catcode `\\12\catcode `\$12\catcode `\&12\catcode `\#12\catcode `\^12\catcode `\_12\catcode `\%12\relax}%
\providecommand \@@startlink[1]{}%
\providecommand \@@endlink[0]{}%
\providecommand \url  [0]{\begingroup\@sanitize@url \@url }%
\providecommand \@url [1]{\endgroup\@href {#1}{\urlprefix }}%
\providecommand \urlprefix  [0]{URL }%
\providecommand \Eprint [0]{\href }%
\providecommand \doibase [0]{https://doi.org/}%
\providecommand \selectlanguage [0]{\@gobble}%
\providecommand \bibinfo  [0]{\@secondoftwo}%
\providecommand \bibfield  [0]{\@secondoftwo}%
\providecommand \translation [1]{[#1]}%
\providecommand \BibitemOpen [0]{}%
\providecommand \bibitemStop [0]{}%
\providecommand \bibitemNoStop [0]{.\EOS\space}%
\providecommand \EOS [0]{\spacefactor3000\relax}%
\providecommand \BibitemShut  [1]{\csname bibitem#1\endcsname}%
\let\auto@bib@innerbib\@empty
\bibitem [{\citenamefont {Bertone}\ \emph {et~al.}(2005)\citenamefont {Bertone}, \citenamefont {Hooper},\ and\ \citenamefont {Silk}}]{Bertone:2004pz}%
  \BibitemOpen
  \bibfield  {author} {\bibinfo {author} {\bibfnamefont {G.}~\bibnamefont {Bertone}}, \bibinfo {author} {\bibfnamefont {D.}~\bibnamefont {Hooper}},\ and\ \bibinfo {author} {\bibfnamefont {J.}~\bibnamefont {Silk}},\ }\href {https://doi.org/10.1016/j.physrep.2004.08.031} {\bibfield  {journal} {\bibinfo  {journal} {Phys. Rept.}\ }\textbf {\bibinfo {volume} {405}},\ \bibinfo {pages} {279} (\bibinfo {year} {2005})},\ \Eprint {https://arxiv.org/abs/hep-ph/0404175} {arXiv:hep-ph/0404175} \BibitemShut {NoStop}%
\bibitem [{\citenamefont {Peccei}\ and\ \citenamefont {Quinn}(1977{\natexlab{a}})}]{Peccei:1977hh}%
  \BibitemOpen
  \bibfield  {author} {\bibinfo {author} {\bibfnamefont {R.~D.}\ \bibnamefont {Peccei}}\ and\ \bibinfo {author} {\bibfnamefont {H.~R.}\ \bibnamefont {Quinn}},\ }\href {https://doi.org/10.1103/PhysRevLett.38.1440} {\bibfield  {journal} {\bibinfo  {journal} {Phys. Rev. Lett.}\ }\textbf {\bibinfo {volume} {38}},\ \bibinfo {pages} {1440} (\bibinfo {year} {1977}{\natexlab{a}})}\BibitemShut {NoStop}%
\bibitem [{\citenamefont {Peccei}\ and\ \citenamefont {Quinn}(1977{\natexlab{b}})}]{Peccei:1977ur}%
  \BibitemOpen
  \bibfield  {author} {\bibinfo {author} {\bibfnamefont {R.~D.}\ \bibnamefont {Peccei}}\ and\ \bibinfo {author} {\bibfnamefont {H.~R.}\ \bibnamefont {Quinn}},\ }\href {https://doi.org/10.1103/PhysRevD.16.1791} {\bibfield  {journal} {\bibinfo  {journal} {Phys. Rev. D}\ }\textbf {\bibinfo {volume} {16}},\ \bibinfo {pages} {1791} (\bibinfo {year} {1977}{\natexlab{b}})}\BibitemShut {NoStop}%
\bibitem [{\citenamefont {Dine}\ and\ \citenamefont {Fischler}(1983)}]{Dine:1982ah}%
  \BibitemOpen
  \bibfield  {author} {\bibinfo {author} {\bibfnamefont {M.}~\bibnamefont {Dine}}\ and\ \bibinfo {author} {\bibfnamefont {W.}~\bibnamefont {Fischler}},\ }\href {https://doi.org/10.1016/0370-2693(83)90639-1} {\bibfield  {journal} {\bibinfo  {journal} {Phys. Lett. B}\ }\textbf {\bibinfo {volume} {120}},\ \bibinfo {pages} {137} (\bibinfo {year} {1983})}\BibitemShut {NoStop}%
\bibitem [{\citenamefont {Abbott}\ and\ \citenamefont {Sikivie}(1983)}]{Abbott:1982af}%
  \BibitemOpen
  \bibfield  {author} {\bibinfo {author} {\bibfnamefont {L.~F.}\ \bibnamefont {Abbott}}\ and\ \bibinfo {author} {\bibfnamefont {P.}~\bibnamefont {Sikivie}},\ }\href {https://doi.org/10.1016/0370-2693(83)90638-X} {\bibfield  {journal} {\bibinfo  {journal} {Phys. Lett. B}\ }\textbf {\bibinfo {volume} {120}},\ \bibinfo {pages} {133} (\bibinfo {year} {1983})}\BibitemShut {NoStop}%
\bibitem [{\citenamefont {Preskill}\ \emph {et~al.}(1983)\citenamefont {Preskill}, \citenamefont {Wise},\ and\ \citenamefont {Wilczek}}]{Preskill:1982cy}%
  \BibitemOpen
  \bibfield  {author} {\bibinfo {author} {\bibfnamefont {J.}~\bibnamefont {Preskill}}, \bibinfo {author} {\bibfnamefont {M.~B.}\ \bibnamefont {Wise}},\ and\ \bibinfo {author} {\bibfnamefont {F.}~\bibnamefont {Wilczek}},\ }\href {https://doi.org/10.1016/0370-2693(83)90637-8} {\bibfield  {journal} {\bibinfo  {journal} {Phys. Lett. B}\ }\textbf {\bibinfo {volume} {120}},\ \bibinfo {pages} {127} (\bibinfo {year} {1983})}\BibitemShut {NoStop}%
\bibitem [{\citenamefont {Andriamonje}\ \emph {et~al.}(2007)\citenamefont {Andriamonje} \emph {et~al.}}]{CAST:2007jps}%
  \BibitemOpen
  \bibfield  {author} {\bibinfo {author} {\bibfnamefont {S.}~\bibnamefont {Andriamonje}} \emph {et~al.} (\bibinfo {collaboration} {CAST}),\ }\href {https://doi.org/10.1088/1475-7516/2007/04/010} {\bibfield  {journal} {\bibinfo  {journal} {J. Cosmol. Astropart. Phys.}\ }\textbf {\bibinfo {volume} {04}},\ \bibinfo {pages} {010}},\ \Eprint {https://arxiv.org/abs/hep-ex/0702006} {arXiv:hep-ex/0702006} \BibitemShut {NoStop}%
\bibitem [{\citenamefont {Anastassopoulos}\ \emph {et~al.}(2017)\citenamefont {Anastassopoulos}, \citenamefont {others},\ and\ \citenamefont {{CAST Collaboration}}}]{CAST:2017uph}%
  \BibitemOpen
  \bibfield  {author} {\bibinfo {author} {\bibfnamefont {V.}~\bibnamefont {Anastassopoulos}}, \bibinfo {author} {\bibnamefont {others}},\ and\ \bibinfo {author} {\bibnamefont {{CAST Collaboration}}} (\bibinfo {collaboration} {CAST collaboration}),\ }\href {https://doi.org/10.1038/nphys4109} {\bibfield  {journal} {\bibinfo  {journal} {Nat. Phys.}\ }\textbf {\bibinfo {volume} {13}},\ \bibinfo {pages} {584} (\bibinfo {year} {2017})}\BibitemShut {NoStop}%
\bibitem [{\citenamefont {Ahmed}\ \emph {et~al.}(2009)\citenamefont {Ahmed} \emph {et~al.}}]{CDMS:2009fba}%
  \BibitemOpen
  \bibfield  {author} {\bibinfo {author} {\bibfnamefont {Z.}~\bibnamefont {Ahmed}} \emph {et~al.},\ }\href {https://doi.org/10.1103/PhysRevLett.103.141802} {\bibfield  {journal} {\bibinfo  {journal} {Phys. Rev. Lett.}\ }\textbf {\bibinfo {volume} {103}},\ \bibinfo {pages} {141802} (\bibinfo {year} {2009})}\BibitemShut {NoStop}%
\bibitem [{\citenamefont {Albakry}\ \emph {et~al.}(2023)\citenamefont {Albakry} \emph {et~al.}}]{SuperCDMS:2022kse}%
  \BibitemOpen
  \bibfield  {author} {\bibinfo {author} {\bibfnamefont {M.~F.}\ \bibnamefont {Albakry}} \emph {et~al.} (\bibinfo {collaboration} {SuperCDMS}),\ }\href {https://doi.org/10.48550/arXiv.2203.08463} {\bibinfo {title} {A {{Strategy}} for {{Low-Mass Dark Matter Searches}} with {{Cryogenic Detectors}} in the {{SuperCDMS SNOLAB Facility}}}} (\bibinfo {year} {2023}),\ \Eprint {https://arxiv.org/abs/2203.08463} {arXiv:2203.08463} \BibitemShut {NoStop}%
\bibitem [{\citenamefont {Creswick}\ \emph {et~al.}(1998)\citenamefont {Creswick}, \citenamefont {Avignone}, \citenamefont {Farach}, \citenamefont {Collar}, \citenamefont {Gattone}, \citenamefont {Nussinov},\ and\ \citenamefont {Zioutas}}]{Creswick:1997pg}%
  \BibitemOpen
  \bibfield  {author} {\bibinfo {author} {\bibfnamefont {R.~J.}\ \bibnamefont {Creswick}}, \bibinfo {author} {\bibfnamefont {F.~T.}\ \bibnamefont {Avignone}, \bibfnamefont {III}}, \bibinfo {author} {\bibfnamefont {H.~A.}\ \bibnamefont {Farach}}, \bibinfo {author} {\bibfnamefont {J.~I.}\ \bibnamefont {Collar}}, \bibinfo {author} {\bibfnamefont {A.~O.}\ \bibnamefont {Gattone}}, \bibinfo {author} {\bibfnamefont {S.}~\bibnamefont {Nussinov}},\ and\ \bibinfo {author} {\bibfnamefont {K.}~\bibnamefont {Zioutas}},\ }\href {https://doi.org/10.1016/S0370-2693(98)00183-X} {\bibfield  {journal} {\bibinfo  {journal} {Phys. Lett. B}\ }\textbf {\bibinfo {volume} {427}},\ \bibinfo {pages} {235} (\bibinfo {year} {1998})},\ \Eprint {https://arxiv.org/abs/hep-ph/9708210} {arXiv:hep-ph/9708210} \BibitemShut {NoStop}%
\bibitem [{\citenamefont {Ballou}\ \emph {et~al.}(2015)\citenamefont {Ballou} \emph {et~al.}}]{OSQAR:2015qdv}%
  \BibitemOpen
  \bibfield  {author} {\bibinfo {author} {\bibfnamefont {R.}~\bibnamefont {Ballou}} \emph {et~al.} (\bibinfo {collaboration} {OSQAR Collaboration}),\ }\href {https://doi.org/10.1103/PhysRevD.92.092002} {\bibfield  {journal} {\bibinfo  {journal} {Phys. Rev. D}\ }\textbf {\bibinfo {volume} {92}},\ \bibinfo {pages} {092002} (\bibinfo {year} {2015})}\BibitemShut {NoStop}%
\bibitem [{\citenamefont {Kozlowski}\ \emph {et~al.}(2024)\citenamefont {Kozlowski} \emph {et~al.}}]{Kozlowski:2024jzm}%
  \BibitemOpen
  \bibfield  {author} {\bibinfo {author} {\bibfnamefont {T.}~\bibnamefont {Kozlowski}} \emph {et~al.},\ }\href@noop {} {\bibinfo {title} {Design and {{Performance}} of the {{ALPS II Regeneration Cavity}}}} (\bibinfo {year} {2024}),\ \Eprint {https://arxiv.org/abs/2408.13218} {arXiv:2408.13218} \BibitemShut {NoStop}%
\bibitem [{\citenamefont {Ehret}\ \emph {et~al.}(2010)\citenamefont {Ehret} \emph {et~al.}}]{Ehret:2010mh}%
  \BibitemOpen
  \bibfield  {author} {\bibinfo {author} {\bibfnamefont {K.}~\bibnamefont {Ehret}} \emph {et~al.},\ }\href {https://doi.org/10.1016/j.physletb.2010.04.066} {\bibfield  {journal} {\bibinfo  {journal} {Phys. Lett. B}\ }\textbf {\bibinfo {volume} {689}},\ \bibinfo {pages} {149} (\bibinfo {year} {2010})}\BibitemShut {NoStop}%
\bibitem [{\citenamefont {Della~Valle}\ \emph {et~al.}(2016)\citenamefont {Della~Valle} \emph {et~al.}}]{DellaValle:2015xxa}%
  \BibitemOpen
  \bibfield  {author} {\bibinfo {author} {\bibfnamefont {F.}~\bibnamefont {Della~Valle}} \emph {et~al.},\ }\href {https://doi.org/10.1140/epjc/s10052-015-3869-8} {\bibfield  {journal} {\bibinfo  {journal} {Eur. Phys. J. C}\ }\textbf {\bibinfo {volume} {76}},\ \bibinfo {pages} {24} (\bibinfo {year} {2016})}\BibitemShut {NoStop}%
\bibitem [{\citenamefont {Ejlli}\ \emph {et~al.}(2020)\citenamefont {Ejlli} \emph {et~al.}}]{Ejlli:2020yhk}%
  \BibitemOpen
  \bibfield  {author} {\bibinfo {author} {\bibfnamefont {A.}~\bibnamefont {Ejlli}} \emph {et~al.},\ }\href {https://doi.org/10.1016/j.physrep.2020.06.001} {\bibfield  {journal} {\bibinfo  {journal} {Phys. Rep.}\ }\textbf {\bibinfo {volume} {871}},\ \bibinfo {pages} {1} (\bibinfo {year} {2020})}\BibitemShut {NoStop}%
\bibitem [{\citenamefont {Buchmuller}\ and\ \citenamefont {Hoogeveen}(1990)}]{Buchmuller:1989rb}%
  \BibitemOpen
  \bibfield  {author} {\bibinfo {author} {\bibfnamefont {W.}~\bibnamefont {Buchmuller}}\ and\ \bibinfo {author} {\bibfnamefont {F.}~\bibnamefont {Hoogeveen}},\ }\href {https://doi.org/10.1016/0370-2693(90)91444-G} {\bibfield  {journal} {\bibinfo  {journal} {Phys. Lett. B}\ }\textbf {\bibinfo {volume} {237}},\ \bibinfo {pages} {278} (\bibinfo {year} {1990})}\BibitemShut {NoStop}%
\bibitem [{\citenamefont {Henke}\ \emph {et~al.}(1993)\citenamefont {Henke}, \citenamefont {Gullikson},\ and\ \citenamefont {Davis}}]{Henke:1993eda}%
  \BibitemOpen
  \bibfield  {author} {\bibinfo {author} {\bibfnamefont {B.~L.}\ \bibnamefont {Henke}}, \bibinfo {author} {\bibfnamefont {E.~M.}\ \bibnamefont {Gullikson}},\ and\ \bibinfo {author} {\bibfnamefont {J.~C.}\ \bibnamefont {Davis}},\ }\href {https://doi.org/10.1006/adnd.1993.1013} {\bibfield  {journal} {\bibinfo  {journal} {Atom. Data Nucl. Data Tabl.}\ }\textbf {\bibinfo {volume} {54}},\ \bibinfo {pages} {181} (\bibinfo {year} {1993})}\BibitemShut {NoStop}%
\bibitem [{\citenamefont {Yamaji}\ \emph {et~al.}(2017)\citenamefont {Yamaji}, \citenamefont {Yamazaki}, \citenamefont {Tamasaku},\ and\ \citenamefont {Namba}}]{Yamaji:2017pep}%
  \BibitemOpen
  \bibfield  {author} {\bibinfo {author} {\bibfnamefont {T.}~\bibnamefont {Yamaji}}, \bibinfo {author} {\bibfnamefont {T.}~\bibnamefont {Yamazaki}}, \bibinfo {author} {\bibfnamefont {K.}~\bibnamefont {Tamasaku}},\ and\ \bibinfo {author} {\bibfnamefont {T.}~\bibnamefont {Namba}},\ }\href {https://doi.org/10.1103/PhysRevD.96.115001} {\bibfield  {journal} {\bibinfo  {journal} {Phys. Rev. D}\ }\textbf {\bibinfo {volume} {96}},\ \bibinfo {pages} {115001} (\bibinfo {year} {2017})},\ \Eprint {https://arxiv.org/abs/1709.03299} {arXiv:1709.03299} \BibitemShut {NoStop}%
\bibitem [{\citenamefont {Halliday}\ \emph {et~al.}(2025)\citenamefont {Halliday} \emph {et~al.}}]{Halliday:2024lca}%
  \BibitemOpen
  \bibfield  {author} {\bibinfo {author} {\bibfnamefont {J.~W.~D.}\ \bibnamefont {Halliday}} \emph {et~al.},\ }\href {https://doi.org/10.1103/PhysRevLett.134.055001} {\bibfield  {journal} {\bibinfo  {journal} {Phys. Lett. Lett.}\ }\textbf {\bibinfo {volume} {134}},\ \bibinfo {pages} {055001} (\bibinfo {year} {2025})},\ \Eprint {https://arxiv.org/abs/2404.17333} {arXiv:2404.17333} \BibitemShut {NoStop}%
\bibitem [{\citenamefont {Peskin}(1995)}]{Peskin:QFT}%
  \BibitemOpen
  \bibfield  {author} {\bibinfo {author} {\bibfnamefont {M.~E.}\ \bibnamefont {Peskin}},\ }\href@noop {} {\emph {\bibinfo {title} {An {{Introduction}} to {{Quantum Field Theory}}}}}\ (\bibinfo  {publisher} {Westview Press},\ \bibinfo {year} {1995})\BibitemShut {NoStop}%
\bibitem [{CaF(2024)}]{CaF2:2024Wikipedia:nov}%
  \BibitemOpen
  \href@noop {} {\bibinfo {title} {Calcium fluoride}},\ \bibinfo {howpublished} {\url{https://en.wikipedia.org/w/index.php?title=Calcium\_fluoride\&oldid=1256066213}} (\bibinfo {year} {2024})\BibitemShut {NoStop}%
\bibitem [{\citenamefont {Polyanskiy}(2024)}]{Polyanskiy:RefractiveIndex}%
  \BibitemOpen
  \bibfield  {author} {\bibinfo {author} {\bibfnamefont {M.~N.}\ \bibnamefont {Polyanskiy}},\ }\href {https://doi.org/10.1038/s41597-023-02898-2} {\bibfield  {journal} {\bibinfo  {journal} {Sci. Data}\ }\textbf {\bibinfo {volume} {11}},\ \bibinfo {pages} {94} (\bibinfo {year} {2024})}\BibitemShut {NoStop}%
\bibitem [{ALP()}]{ALPSII}%
  \BibitemOpen
  \href@noop {} {\bibinfo {title} {{{ALPS II}}}},\ \bibinfo {howpublished} {\url{https://alps.desy.de/our\_activities/axion\_wisp\_experiments/alps\_ii/}}\BibitemShut {NoStop}%
\bibitem [{\citenamefont {O'Hare}(2020)}]{AxionLimits}%
  \BibitemOpen
  \bibfield  {author} {\bibinfo {author} {\bibfnamefont {C.}~\bibnamefont {O'Hare}},\ }\href {https://doi.org/10.5281/zenodo.3932430} {\bibinfo {title} {Cajohare/{{AxionLimits}}: {{AxionLimits}}}} (\bibinfo {year} {2020})\BibitemShut {NoStop}%
\bibitem [{\citenamefont {Matsumoto}\ \emph {et~al.}(2024)\citenamefont {Matsumoto}, \citenamefont {Sheng},\ and\ \citenamefont {Xing}}]{Matsumoto:2024fzr}%
  \BibitemOpen
  \bibfield  {author} {\bibinfo {author} {\bibfnamefont {S.}~\bibnamefont {Matsumoto}}, \bibinfo {author} {\bibfnamefont {J.}~\bibnamefont {Sheng}},\ and\ \bibinfo {author} {\bibfnamefont {C.-Y.}\ \bibnamefont {Xing}},\ }\href {https://doi.org/10.48550/arXiv.2409.09950} {\bibinfo {title} {Detection of {{Dark Matter Coherent Scattering}} via {{Torsion Balance}} with {{Test Bodies}} of {{Different Sizes}}}} (\bibinfo {year} {2024}),\ \Eprint {https://arxiv.org/abs/2409.09950} {arXiv:2409.09950 [hep-ph]} \BibitemShut {NoStop}%
\bibitem [{Hel()}]{HeliosSputteringTool}%
  \BibitemOpen
  \href@noop {} {\bibinfo {title} {Helios {{Sputtering Tool}} {\textbar} {{B{\"u}hler Leybold Optics}} {\textbar} {{B{\"u}hler Group}}}},\ \bibinfo {howpublished} {\url{https://www.buhlergroup.cn/global/en/products/leybold\_optics\_heliosseriesprecisionopticsvacuumcoater.html}}\BibitemShut {NoStop}%
\bibitem [{\citenamefont {Baddour}(2011)}]{Baddour:2011:2dFourier}%
  \BibitemOpen
  \bibfield  {author} {\bibinfo {author} {\bibfnamefont {N.}~\bibnamefont {Baddour}},\ }in\ \href {https://doi.org/10.1016/B978-0-12-385861-0.00001-4} {\emph {\bibinfo {booktitle} {Advances in Imaging and Electron Physics}}},\ \bibinfo {series} {Advances in Imaging and Electron Physics}, Vol.\ \bibinfo {volume} {165},\ \bibinfo {editor} {edited by\ \bibinfo {editor} {\bibfnamefont {P.~W.}\ \bibnamefont {Hawkes}}}\ (\bibinfo  {publisher} {Elsevier},\ \bibinfo {year} {2011})\ pp.\ \bibinfo {pages} {1--45}\BibitemShut {NoStop}%
\bibitem [{\citenamefont {An}\ \emph {et~al.}(2024)\citenamefont {An}, \citenamefont {Chen}, \citenamefont {Liu}, \citenamefont {Sheng},\ and\ \citenamefont {Zhang}}]{An:2024kns}%
  \BibitemOpen
  \bibfield  {author} {\bibinfo {author} {\bibfnamefont {X.}~\bibnamefont {An}}, \bibinfo {author} {\bibfnamefont {M.}~\bibnamefont {Chen}}, \bibinfo {author} {\bibfnamefont {J.}~\bibnamefont {Liu}}, \bibinfo {author} {\bibfnamefont {Z.}~\bibnamefont {Sheng}},\ and\ \bibinfo {author} {\bibfnamefont {J.}~\bibnamefont {Zhang}},\ }\href {https://doi.org/10.1063/5.0226159} {\bibfield  {journal} {\bibinfo  {journal} {Matter Radiat. Extremes}\ }\textbf {\bibinfo {volume} {9}},\ \bibinfo {pages} {067204} (\bibinfo {year} {2024})},\ \Eprint {https://arxiv.org/abs/2406.16796} {arXiv:2406.16796} \BibitemShut {NoStop}%
\bibitem [{\citenamefont {Jackson}(1999)}]{Jackson1999:ClassicalElectrodynamics}%
  \BibitemOpen
  \bibfield  {author} {\bibinfo {author} {\bibfnamefont {J.~D.}\ \bibnamefont {Jackson}},\ }\href@noop {} {\emph {\bibinfo {title} {Classical {{Electrodynamics}}}}},\ \bibinfo {edition} {3rd}\ ed.\ (\bibinfo  {publisher} {JOHN WILEY \& SONS, INC.},\ \bibinfo {year} {1999})\BibitemShut {NoStop}%
\end{thebibliography}
\end{document}